\documentclass[preprint2]{proto}
\usepackage{natbib}
\bibpunct{(}{)}{;}{a}{,}{,}

\newcommand*{\sign}{\mathop{\mathrm{sign}}\nolimits}
\newcommand{\bmid}{\beta_\mathrm{mid}}
\newcommand{\ppd}{protoplanetary disk}
\begin{document}
\title{\textbf{\LARGE Transport and Accretion in Planet-Forming Disks}}
\author{
  \textbf{Neal~J.~Turner}\\
  \textit{Jet Propulsion Laboratory, California Institute of Technology}\\[3mm]
  \textbf{S\'ebastien~Fromang}\\
  \textit{Service d'Astrophysique de Saclay}\\[3mm]
  \textbf{Charles~Gammie}\\
  \textit{University of Illinois at Urbana-Champaign}\\[3mm]
  \textbf{Hubert~Klahr}\\
  \textit{Max Planck Institute for Astronomy}\\[3mm]
  \textbf{Geoffroy~Lesur}\\
  \textit{Institute for Planetology and Astrophysics of Grenoble}\\[3mm]
  \textbf{Mark~Wardle}\\
  \textit{Macquarie University}\\[3mm]
  \textbf{Xue-Ning~Bai}\\
  \textit{Harvard-Smithsonian Center for Astrophysics}}
\begin{abstract}
\baselineskip = 11pt
\leftskip = 0.65in 
\rightskip = 0.65in
\parindent=1pc

\noindent {\small Planets appear to form in environments shaped by the
  gas flowing through protostellar disks to the central young stars.
  The flows in turn are governed by orbital angular momentum transfer.
  In this chapter we summarize current understanding of the transfer
  processes best able to account for the flows, including
  magneto-rotational turbulence, magnetically-launched winds,
  self-gravitational instability and vortices driven by hydrodynamical
  instabilities.  For each in turn we outline the major achievements
  of the past few years and the outstanding questions.  We underscore
  the requirements for operation, especially ionization for the
  magnetic processes and heating and cooling for the others.  We
  describe the distribution and strength of the resulting flows and
  compare with the long-used phenomenological $\alpha$-picture,
  highlighting issues where the fuller physical picture yields
  substantially different answers.  We also discuss the links between
  magnetized turbulence and magnetically-launched outflows, and
  between magnetized turbulence and hydrodynamical vortices.  We end
  with a summary of the status of efforts to detect specific
  signatures of the flows.}  \\~\\~\\~
%leave this in to get the correct vertical space after the abstract
\end{abstract}

%%%%%%%%%%%%%%%%%%%%%%%%%%%%%%%%%%%%%%%%%%%%%%%%%%%%%%%%%%%%%%%%%%%%%%%%
\section{\textbf{Introduction}}
A key to the evolution of the planet-forming material in protostellar
disks is the angular momentum.  The angular momentum per unit mass of
gas orbiting just above the star's surface is only 1\% that of gas
near the disk's outer edge.  Sustaining the accretion on the star
therefore requires taking almost all the angular momentum out of the
accreting matter.  Since the angular momentum cannot be destroyed it
must be handed off to other material.  The gas receiving the angular
momentum could move radially outward, or could join outflows escaping
above and below the disk.  The gas losing angular momentum and
spiraling inward releases gravitational potential energy into
turbulence, vortices or magnetic dissipation, which can stir and heat
the flow.

Angular momentum transport processes that could be important under the
right circumstances include magnetohydrodynamical turbulence initiated
by the magneto-rotational instability, outflows, hydrodynamical
processes and gravitational instability.  In several of these the
large-scale transport is built up from smaller-scale gas motions such
as turbulent eddies, spiral waves or long-lived vortices.  In addition
to governing the flow of gas to the star, these small-scale motions
can control the mixing of gas molecules and dust grains along the
radial and vertical gradients in temperature, density and radiation
intensity \citep{2006a&a...452..751f, 2006Icar..181..178C,
  2010ApJ...708..188T, 2011apjs..196...25s}.  The small-scale motions
also can alter the growth of dust grains into bigger bodies
\citep{2008a&a...480..859b, 2009ApJ...698.1122O, 2010a&a...513a..57z},
the collisional evolution of planetesimals \citep{2011MNRAS.415.3291G}
and the orbital migration of protoplanets \citep{2004MNRAS.350..849N}.

In this chapter we review what is known about the basic operation of
these transport processes and when and in what regions they work.  We
restrict ourselves to processes originating in the disk.  For example
we do not consider either the influence of a companion star
\citep{1993prpl.conf..749L}, the shear between the disk surface and
infalling material \citep{1982MitAG..57...49R} or the interaction of
the disk with the young star's magnetosphere
\citep{2013MNRAS.430..699R}.  Furthermore we devote most attention to
work published since the previous volume in this series.

We examine the magneto-rotational instability's linear growth
(\S\ref{sec:linearmri}) and development into saturated turbulence
(\S\ref{sec:saturation}); disk-driven winds (\S\ref{sec:winds}); and
self-gravitational (\S\ref{sec:GI}) and hydrodynamical instabilities
(\S\ref{sec:weather}).  Finally we discuss the prospects for observing
the signatures of each transport process (\S\ref{sec:signatures}) and
the outlook for the next few years (\S\ref{sec:outlook}).

%%%%%%%%%%%%%%%%%%%%%%%%%%%%%%%%%%%%%%%%%%%%%%%%%%%%%%%%%%%%%%%%%%%%%%%%
\section{\textbf{The Linear Magneto-Rotational Instability}
  \label{sec:linearmri}}
Study of the linear magneto-rotational instability (MRI) is warranted
by the desire to understand the mechanism driving
magnetohydrodynamical (MHD) turbulence and the conditions under which
driving occurs, so we can map the MRI-active regions in \ppd s.  The
disks' weak ionization can lead to magnetic fluctuations being damped.
While linear analysis does not address the non-linear saturated state,
it provides a good check on numerical codes during the initial
departure from equilibrium.  Local linear analyses have also proven
surprisingly good at pinpointing the boundaries of the MHD turbulent
regions in stratified simulations.

We first review the behavior of the MRI in ideal MHD and then describe
the different behaviors in the presence of Ohmic and ambipolar
diffusion and Hall drift.  We conclude with a discussion of the
ionization equilibrium and the implications for magnetic activity.

\subsection{\textbf{MRI Under Ideal MHD}}

MRI in ideal MHD Couette flows was discovered in the late 1950s
\citep{1960PNAS...46..253C,V59} but its astrophysical applications
were appreciated only with rediscovery in the context of accretion
disks \citep{1991ApJ...376..214B}.  The basic mechanism relies on
angular momentum transfer by magnetic tension coupling fluid elements
at neighbouring radii.  As the fluid elements drift apart under the
orbital shear, magnetic tension transfers angular momentum from the
inner fluid element to the outer one.  The inner element moves inward
and speeds up, the outer moves outward and slows down, increasing the
rate at which the elements drift apart and leading to a runaway.  The
key is that the magnetic field is weak enough so magnetic tension
cannot overcome the tendency of the fluid elements to separate due to
the shear.

The result is a robust instability, relying only on a magnetic field
embedded in a rotating shear flow in which angular momentum decreases
outward.  For a simple initially-vertical magnetic field, the most
unstable mode is axisymmetric with growth rate $\frac{3}{4}\Omega$
\citep{1991ApJ...376..214B} and vertical wavelength near $2\pi
v_{Az}/\Omega$.  This rate has proven to be a general maximum in
collisional plasmas under Keplerian rotation.  The growth rate is
reduced if the plasma beta parameter
$\beta=P_\mathrm{gas}/P_\mathrm{magnetic}<1$, because the
fastest-growing mode no longer fits within the disk thickness.  Adding
a toroidal magnetic component such that $\beta<1$ reduces the fastest
linear growth rate without much changing the wavelength.  The growth
rate approaches zero as $\beta\rightarrow 0$ if the field is within
$30^\circ$ of toroidal \citep{2000ApJ...540..372K}.  For strictly
toroidal fields, the fastest modes' azimuthal wavelength again is near
$2\pi v_{A\phi}/\Omega$ but the vertical wavelength is vanishingly
small \citep{1996MNRAS.279..767T}.  As with inclined fields, growth is
slower if $\beta\la 1$ \citep{2000ApJ...540..372K}.  When the toroidal
fields' $\beta\lesssim 10$, the Parker magnetic buoyancy instability
grows faster than the MRI \citep{1996MNRAS.279..767T} and expels some
of the magnetic energy into the disk atmosphere
\citep{hirose&turner11}.

\subsection{\textbf{MRI Under Ohmic MHD}}
The effect of Ohmic resistivity $\eta$ on the MRI is straightforward:
it tends to erase magnetic fluctuations on length scales shorter than
$\eta/v_A$.  The important length scale for the MRI is the fastest
growing wavelength $2\pi v_A/\Omega$, so the dimensionless number that
determines its effect on the MRI is the Elsasser number $\Lambda =
v_A^2/(\eta\Omega)$.  Local and global linear instabilities are damped
when the time to diffuse across the fastest-growing wavelength is less
than the growth time, corresponding to $\Lambda < 1$
\citep{1996ApJ...457..798J, 1999ApJ...515..776S}.  This linear
criterion is also a good estimator for the onset of turbulence
\citep{2001ApJ...561L.179S, 2007ApJ...659..729T}.

\subsection{\textbf{MRI Under Ambipolar MHD}\label{sec:linearMRIwAD}}
MRI under ambipolar diffusion behaves similarly to the Ohmic case but
there are some subtle differences because ambipolar diffusion does not
dissipate field-parallel currents.  For strictly vertical initial
fields, there are no field-parallel currents in the initial and
perturbed states and the linear MRI behaves as in the Ohmic case, but
with $\Lambda$ replaced by $Am=v_A^2/(\eta_A\Omega)$, where $\eta_A$
is the diffusivity due to ambipolar drift \citep{1994ApJ...421..163B,
  1999MNRAS.307..849W}.

This degeneracy is broken if the initial field has toroidal and
vertical components.  Then perturbations with a restricted range of
wavevector orientations may be destabilised by ambipolar diffusion.
This only partly compensates for the lower growth rate associated with
the tilt in the field or with Ohmic resistivity if this is also
present \citep{2004ApJ...608..509D, 2004MNRAS.348..355K}, and so
appears to be of minor consequence.

\subsection{\textbf{MRI Under Hall MHD}}
Hall drift, characterised by transport coefficient $\eta_H$ and
dimensionless number $\Lambda_H = v_A^2/(\eta_H\Omega)$, differs
fundamentally from resistivity and ambipolar diffusion.  It makes
magnetic field lines drift through the gas in the direction of the
current density, giving a tendency to twist; no dissipation is
involved.  Hall drift therefore permits the magnetic field to
restructure in strange ways with no energy penalty.  For example, Hall
drift tends to rotate the MRI-induced buckling of an initially
vertical field clockwise about the initial direction
\citep{1999MNRAS.307..849W}.  If this is parallel to the rotation
axis, Hall drift acts in concert with the Keplerian shear in
generating toroidal from radial field.  The maximum growth rate
remains at $\frac{3}{4}\Omega$ but the most unstable wavelength
becomes longer.  If the initial field and rotation are antiparallel,
Hall drift acts against the Keplerian shear, and completely suppresses
the MRI if $\Lambda_H<0.5$.

Because Hall drift acts directly with or against the Keplerian shear
in generating $B_\phi$ from $B_r$, it strongly modifies the linear MRI
when $\Lambda_H<1$, irrespective of the magnitudes of $\Lambda$ and
$Am$.  For the ``favorable'' case $B_z>0$, if $\Lambda_H <
min(\Lambda,Am)$ the maximum growth rate is $\frac{3}{4}\Omega$ even
if $\Lambda$ and/or $Am$ are small.  Even if $\Lambda$ and/or
$Am<\Lambda_H$, Hall drift mitigates the damping effects of Ohmic or
ambipolar diffusion, increasing the effective Elsasser number by the
factor $\Lambda_H^{-1/2}$ \citep{1999MNRAS.307..849W,
  2012MNRAS.422.2737W}.

A toroidal component in the initial field introduces trigonometric
corrections to the MRI growth rate and wavelength, and additional
mitigation of Ohmic damping by ambipolar diffusion
\citep{2012MNRAS.423..222P}.

Hall drift is able to restructure the field with no associated
dissipation.  Its ability to mitigate or exacerbate damping of the
MRI, and the fact that Ohmic and ambipolar diffusion and Hall drift
all scale differently with the field strength $B$, may give rise to
interesting behavior in the non-linear state.

\subsection{\textbf{Ionization and Recombination}}

The transport coefficients $\eta$, $\eta_A$, and $\eta_H$ depend on
the abundances of the charged particles in the weakly ionized gas in
\ppd s.  The ionization sources in \ppd s are thermal collisions,
stellar ultraviolet (UV) and X-ray photons, interstellar cosmic rays
and the decay of radionuclides.  Collisional ionization is important
at $\ga 1000~K$, temperatures found only within 0.1--1~AU of the star
depending on the mass flow rate, and also high in the disk atmosphere.
Only the elements with the lowest ionization potentials are ionized,
so the ionization fraction is capped at the gas-phase abundance of the
most common alkali metals.

Far-UV continuum photons are absorbed in the disk's uppermost
0.001--0.1~g~cm$^{-2}$ \citep{2007prpl.conf..751B,
  2011ApJ...735....8P} unless absorbed first by an intervening disk
wind emerging from smaller radii \citep{2003MNRAS.342..427F,
  2012A&A...538A...2P}.  Lyman-$\alpha$ photons propagate by resonant
scattering and so may enhance ionization rates somewhat deeper in the
disk \citep{2011ApJ...739...78B}.

Stellar X-rays are absorbed in the uppermost 10~g~cm$^{-2}$
\citep{1999apj...518..848i, mohantyetal13}, yielding unattenuated
ionization rates around $10^{-10}$~s$^{-1}$ at 1~AU for the median
X-ray luminosity of $10^{30}$~erg~s$^{-1}$ measured among young
Solar-mass stars in Orion \citep{2000aj....120.1426g}.

Interstellar cosmic rays are largely screened out by the young star's
wind, according to an extrapolation of the Solar relation between spot
coverage and cosmic ray modulation to young stars' greater spot
covering fractions \citep{2013ApJ...772....5I}.  On the other hand, if
the stellar wind is restricted to the poles and the incoming cosmic
rays interact mostly with a disk wind threaded by poloidal magnetic
fields, the energetic particles are both focused and mirrored by the
fields' pinching toward the equatorial plane.  The first increases and
the second decreases the ionization rate, on balance yielding only
small changes compared with ambient conditions
\citep{2004ApJ...602..528D, 2013ApJ...772....5I}.  If cosmic rays do
reach the disk, then they will penetrate much deeper than X-rays, to
100~g~cm$^{-2}$ \citep{1981PASJ...33..617U, 2009apj...690...69u}, but
yield a lower unattenuated ionization rate of about
$10^{-17}$~s$^{-1}$.

The radionuclides ionize at lower rates, and the decay products are
absorbed inside the solid material if contained in particles bigger
than $\sim$1~mm \citep{2013ApJ...764..104U}.  Most of the radionuclide
ionization comes from $^{26}$Al, if present.  Its half-life of 0.7~Myr
and initial abundance inferred from decay products in primitive
meteorites yield an ionization rate $(7-10)\times 10^{-19}$~s$^{-1}$
\citep{2009apj...690...69u}.

Clearly there is a great deal of variation in the overall ionization
rate with distance from the star and depth in the disk.

The electrons and molecular ions created by ionizations may recombine
directly, or by charge transfer to metal ions followed by radiative
recombination \citep[e.g.][]{UN90, 2006A&A...445..205I}.  The latter
rate is relatively low, making metal ions the dominant positive
species in the gas phase except at the highest ionization levels.  At
low temperatures the metal atoms become adsorbed on grain surfaces,
removing them from the picture.  Dust grains play a major role through
the competitive sticking of electrons and ions from the gas phase and
subsequent recombination on the grains \citep{2000apj...543..486s}.

For a given grain population, the important parameter is
$\zeta/n_\mathrm{H}$, the ratio of the ionization rate per H nucleus
to the number of nuclei per unit volume.  At high ionization levels
most charge resides in the gas phase.  Grains acquire a net negative
charge because of the higher thermal speed of the electrons, but the
fraction of negative charge held by grains is small.  As an electron
is more likely to encounter an ion in the gas phase than a grain, the
majority of recombinations occur in the gas phase.  At intermediate
ionization levels, electrons are more likely to stick to a grain
surface before encountering an ion in the gas phase.  In this regime,
most recombinations occur via the sticking of ions to negatively
charged grains.  At low ionization levels, most grains are uncharged,
and ions and electrons tend to stick to neutral grains before
recombining.  Recombinations occur primarily via the collision of
oppositely charged grains.

\subsection{\textbf{Conductivity}}

Ionization fractions are highest in the disk atmosphere, where the UV
and X-ray photons are absorbed (in the upper 0.01 and 10 g\,cm$^{-2}$,
respectively), and low densities reduce the recombination rates.
Deeper in the disk, the ionization levels plummet due to shielding by
the overlying layers and the increase in recombination rates with
density.

The overall ionization level and the relative abundances of ions,
electrons and charged grains therefore vary strongly with depth, with
concomitant changes to the Ohmic and ambipolar diffusivities and Hall
drift.  The relative importance of the non-ideal terms is determined
by the ratio $n_\mathrm{H}/B$, which controls the degree to which
neutral collisions decouple charged species from the magnetic field.
At successively higher densities, first grains, then ions, and finally
electrons are decoupled.  Roughly speaking, ambipolar diffusion
dominates at low densities on strong magnetic fields, when ions and
electrons remain coupled.  The Ohmic term typically dominates at the
high densities near the midplane and on weak magnetic fields.  The
Hall term dominates over a broad range of intermediate conditions,
when ions are decoupled from the fields by collisions with neutrals,
but electrons are not \citep{2007ap&ss.311...35w,
  2008MNRAS.388.1223S}.

Ohmic diffusivity is always determined by the electron fraction even
when electrons are not the most abundant charged particle.  Ambipolar
diffusivity is typically determined by the ion density, but can be set
by the charged grains if these are small and abundant
\citep{2011ApJ...739...51B}; at low ionization levels charged grains
again play a role.  The Hall drift is more complex, as it is
controlled by the mismatch in the degree to which neutral collisions
decouple positive and negative species from the magnetic field.  At
low densities negatively charged grains are decoupled and ions and
electrons are coupled, implying $\eta_H<0$.  At intermediate densities
ions decouple too and $\eta_H>0$.

\subsection{\textbf{Active Layers and Dead Zone}}
A simple local criterion for MRI turbulence is that the linear
instability has an unstable mode with wavelength shorter than the disk
scale height --- i.e., that Ohmic and ambipolar diffusion are not too
severe.  Hall drift enters by modifying the twisting of the field in
response to the Keplerian shear, and may aid or hinder the growth.  By
way of example, for vertical fields this criterion may be written
\begin{equation}
	\left({1\over\Lambda}+{1\over Am}\right)^2  <  \left( 2 +
             {s\over\Lambda_H} \right)
	\left({3\beta\over 16\pi^2} - {1\over 2} - {s\over\Lambda_H} \right)
\end{equation}
where $s=\sign(B_z)$ \citep[][used $1/k$ rather than $\lambda$ in the
  criterion $kh=1$, so in their equivalent expression $\beta$ has
  coefficient $\frac{3}{4}$]{2012MNRAS.422.2737W}.  This criterion
describes reasonably well whether turbulence develops in simulations
in unstratified boxes in resistive MHD \citep{2001ApJ...561L.179S},
ambipolar diffusion \citep{1998ApJ...501..758H, 2011apj...736..144b}
and Hall drift \citep{2002apj...570..314s, 2002apj...577..534s,
  2013MNRAS.434.2295K}.  The criterion also adequately describes the
height of the boundary in stratified local simulations treating
resistivity \citep{2008ApJ...679L.131T, okuzumi&hirose11}.

Most assessments of the dead zone extent in \ppd s included only Ohmic
resistivity.  However, ambipolar diffusion is more effective in the
surface layers.  The MRI turbulence's upper and outer edges tend to be
set by ambipolar diffusion \citep{2011ApJ...727....2P,
  2011ApJ...739...50B} or by the fields' stiffness at high magnetic
pressure \citep{2000apj...534..398m, 2010ApJ...708..188T}.  Polycyclic
aromatic hydrocarbon grains (PAHs) if found at abundances similar to
those in Herbig Ae/Be disks would limit the turbulence to a thin layer
ionized by the stellar FUV in the disk near the star
\citep{2011ApJ...735....8P, 2011ApJ...727....2P}, but at the lower
densities found further out, PAHs enhance magnetic coupling
\citep{2011ApJ...739...51B}.

\begin{figure}[tb!]
   \centering
   \includegraphics[width=0.9\linewidth]{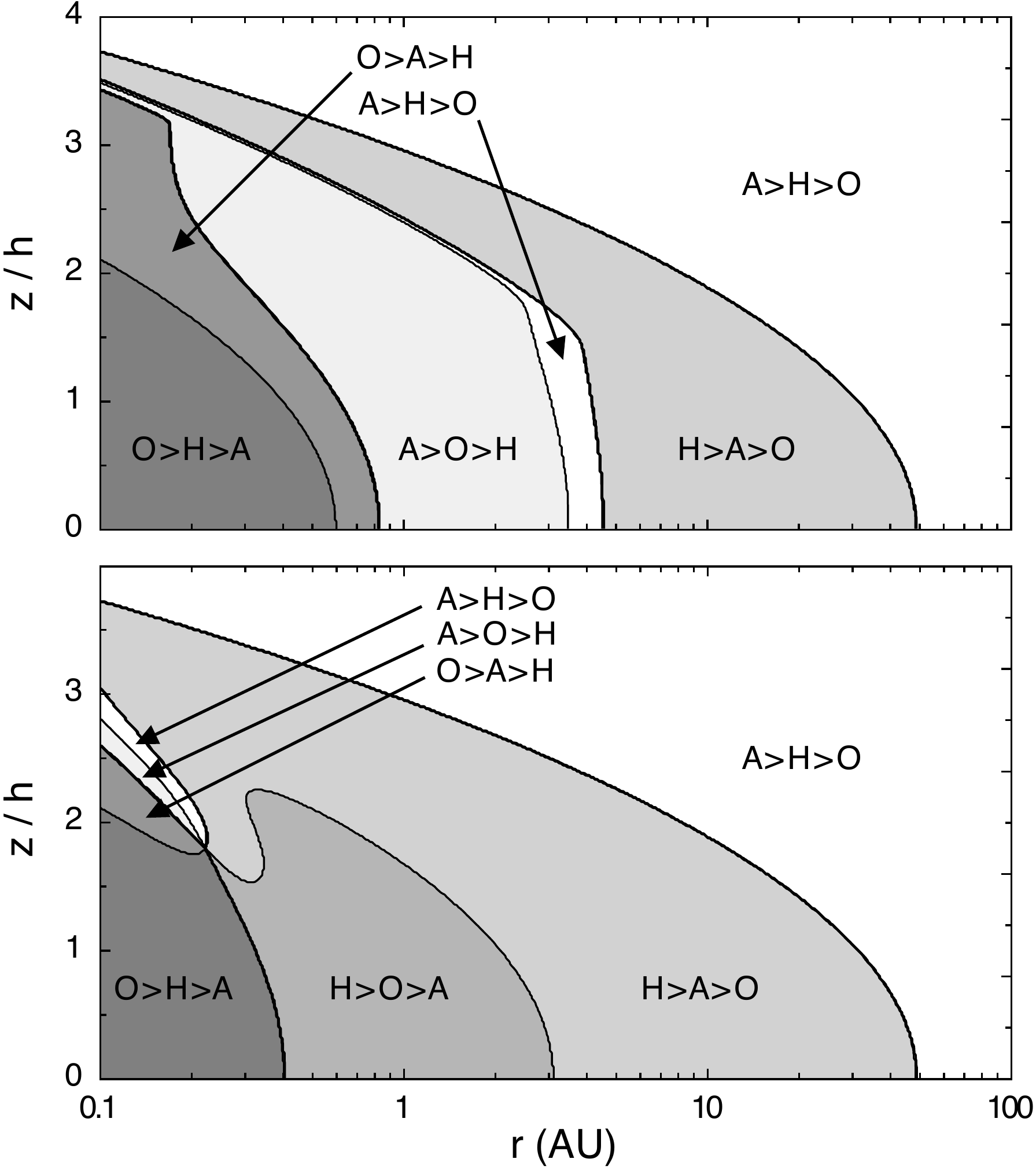}
   \caption{Effect of grains on magnetic diffusion and drift in the
     minimum-mass Solar nebula where $\Sigma=1700
     (r/\mathrm{AU})^{-1.5}$~g\,cm$^{-2}$,
     $T=280\,(r/\mathrm{AU})^{-0.5}$~K and $B$ is constant with height
     and chosen so $v_A=0.1 c_s$ at the midplane.  The axes are the
     radius in AU and height in scale-heights.  Shading indicates the
     ordering of the Ohmic, Hall and ambipolar diffusivities for a
     population of 0.1~$\mu$m grains with 1\% (top panel) or 0.01\%
     (lower panel) of the gas mass, under ionization by X-rays, cosmic
     rays and radioactivity.  \vspace*{-1.5mm}}
   \label{fig:conductivity}
\end{figure}

Hall drift plays a critical role throughout much of the disk
(Fig.~\ref{fig:conductivity}), increasing or decreasing the estimated
active column by orders of magnitude \citep{2012MNRAS.422.2737W}.
However for $\Lambda_H\la5\,v_A/c_s$ the Hall drift leads to most of
the magnetic flux being confined into zonal regions in unstratified
calculations, shutting down turbulent transport
\citep{2013MNRAS.434.2295K}.  This has the potential to greatly reduce
the extent of the MRI turbulent layer.  The following section deals
with the saturation of the MRI and the properties of the resulting
flows.

%%%%%%%%%%%%%%%%%%%%%%%%%%%%%%%%%%%%%%%%%%%%%%%%%%%%%%%%%%%%%%%%%%%%%%%%
\section{\textbf{Saturation of MHD Turbulence}
  \label{sec:saturation}}
While the last section covered the linear MRI, this one focuses on the
non-linear evolution.  By the mid-1990s it was established that the
MRI robustly leads to MHD turbulence transporting angular momentum
outward \citep{1992ApJ...400..595H, 1995ApJ...440..742H,
  1995ApJ...446..741B, hgb96}.  However in recent years, our view of
disk turbulence has been challenged by subtle effects both numerical
and physical, thanks to faster computers and a better understanding of
the microphysics.  The purpose of this section is to review these
findings and discuss the saturation amplitude of the turbulence.  The
amplitude is commonly measured by $\alpha$, the ratio of accretion
stress to gas pressure.  The stress $-B_RB_\phi/4\pi+\rho v_R\delta
v_\phi$ depends on the density $\rho$ and the radial and azimuthal
components of the magnetic field and velocity, the latter measured
relative to the background orbital rotation
\citep{1995ApJ...440..742H}.

\subsection{\textbf{Numerical Approaches}}

Many of the codes now in use employ variations of the conservative
Godunov scheme.  This is the case for example in {\sc Athena}
\citep{stoneetal08}, {\sc Pluto} \citep{mignoneetal07} and {\sc
  Nirvana-III} \citep{ziegler05,ziegler08}.  Alternatives include the
pseudo-spectral code {\sc Snoopy}
(\url{http://ipag.osug.fr/$\sim$glesur/snoopy.html}) and the 6th-order
finite difference {\sc Pencil Code}
(\url{http://pencil-code.nordita.org}), while {\sc Zeus}
\citep{1992apjs...80..753s, 1995CoPhC..89..127H}, which dominated the
early days, is still used by some teams.  With this wealth of methods,
a large variety of setups has been developed for studying the MRI's
non-linear evolution.  Here we have space for only a sampling.

\begin{figure}[tb!]
   \centering
   \includegraphics[width=0.7\linewidth]{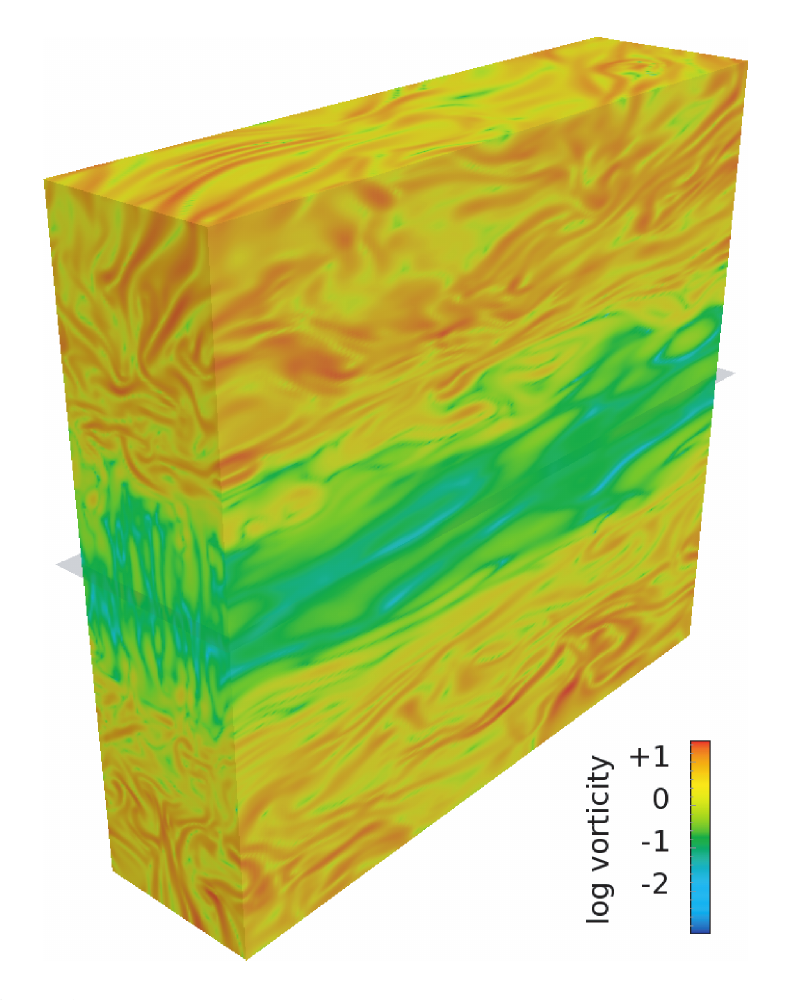}
   \caption{Stratified shearing-box calculation of a small patch of
     protostellar disk.  The Ohmic resistivity is low enough for
     magneto-rotational turbulence only in the X-ray ionized layers on
     the disk's top and bottom surfaces.  Colors indicate the
     vorticity ${\left|\bf \nabla\times v\right|}$ of the velocity
     ${\bf v}$ relative to the Keplerian shear.  The ionized,
     turbulent layers show folded eddy structures, while the flow near
     the midplane is dominated by shearing waves
     \citep{2011MNRAS.415.3291G}.  \vspace*{-1.5mm}}
   \label{fig:shearingbox}
\end{figure}

The most popular of all the approaches is the local shearing box
model.  The differential rotation is linearized in a small Cartesian
volume rotating around the central object.  Compressible and
incompressible formulations are used and the vertical stratification
can be included (Fig.~\ref{fig:shearingbox}).  The shearing box is
useful as a simplified framework for the dynamics.  The problem is
physically well--posed and boundary conditions are straightforward.
High numerical resolutions can be reached and the consequences of
dissipative processes investigated.  We detail recent findings in
\S\S\ref{ideal_sat_sec}--\ref{ambipolar_sat_sec}.  Of course, the
shearing box's advantages come at a cost: curvature terms are ignored
and large-scale gradients are neglected.  These limitations have
motivated the development of global simulations treating the radial
structure.  Spatial resolution is sacrificed, and boundary conditions
require care, but in other respects the realism is better and the
results can be compared directly with observations.  We review recent
global findings in \S\ref{global_sat_sec}.

A key aspect of all simulations is the magnetic field topology.  In
the shearing box this is particularly important, because the magnetic
flux threading the domain is conserved over time due to the boundary
conditions: for example, any net (domain-averaged) vertical field
present initially remains for all time.  Different configurations have
been investigated: vertical and/or toroidal fields yield ``net flux
simulations'' in which the ever-present magnetic flux is a permanent
seed for turbulence.  Configurations with vanishing mean magnetic
field have also been investigated.  In this peculiar situation, the
flow must constantly regenerate its own magnetic field through a
dynamo process.

\subsection{\textbf{Ideal MHD Simulations}
  \label{ideal_sat_sec}}

Here we describe the lessons learned in recent years using local
numerical calculations neglecting all explicit non-ideal magnetic
effects.  To further simplify the problem, the equation of state is
most often taken to be isothermal.  The flow then depends a priori on
the magnetic field topology, box size, grid resolution, and whether
the vertical stratification is included.

The special zero-net-flux or dynamo case has proven to be peculiar.
In the absence of vertical stratification, MRI--driven turbulent
activity decreases as resolution is increased \citep{fromang&pap07,
  guanetal09, simon&hawley09a, bodoetal11}, with no sign of converging
to a well-defined value even at resolutions up to 256~cells per
density scale height.  This is a surprise because when extrapolated to
infinite resolution, it suggests the flow remains laminar in the
absence of a net magnetic flux.  It is still not clear whether this
result is a numerical artifact, as claimed early on
\citep{fromang&pap07} or is due to limitations of the shearing box
\citep{2010A&A...513L...1K, bodoetal11}

Non-convergence, however, appears to be specific to the homogeneous
shearing box without net magnetic field.  When the density
stratification is treated, numerical convergence is reached
\citep{davisetal10, shietal10}.  It is unclear why this differs from
the homogeneous case.  A suggestion is that the episodic escape of
toroidal magnetic fields revealed in the so--called butterfly diagram
\citep{2000apj...534..398m, shietal10, hirose&turner11} mediates a
mean-field dynamo \citep{gressel10}.  Care is needed because the grid
resolutions in stratified simulations are limited, owing to the high
Alfv\'en speeds in the disk corona which necessitate timesteps shorter
than in the homogeneous case.  The lack of convergence of turbulent
transport also disappears when a net magnetic flux, toroidal
\citep{guanetal09} or vertical \citep{simon&hawley09a}, threads the
computational domain.  Again the reason for the difference compared
with zero-net-flux boxes has not been identified, even if it seems
natural to relate it to the linear MRI modes in that case.  With a net
vertical magnetic field, the turbulent activity depends on the box's
radial extent \citep{2008A&A...487....1B} and this comes from the
linear MRI modes and how they are destabilized by parasitic
instabilities.

Clarifying these issues will be tremendously valuable.  However,
regardless of its origins, the convergence issue highlighted in this
section teaches us that the grid resolution in current numerical
simulations affects the outcome in some cases, and thus dissipation
should be treated explicitly.  This is the topic of the following
sections.

\subsection{\textbf{MRI with Ohmic Resistivity}
  \label{ohmic_sat_sec}}

\begin{figure}[tb!]
\centering
\includegraphics[width=0.9\linewidth]{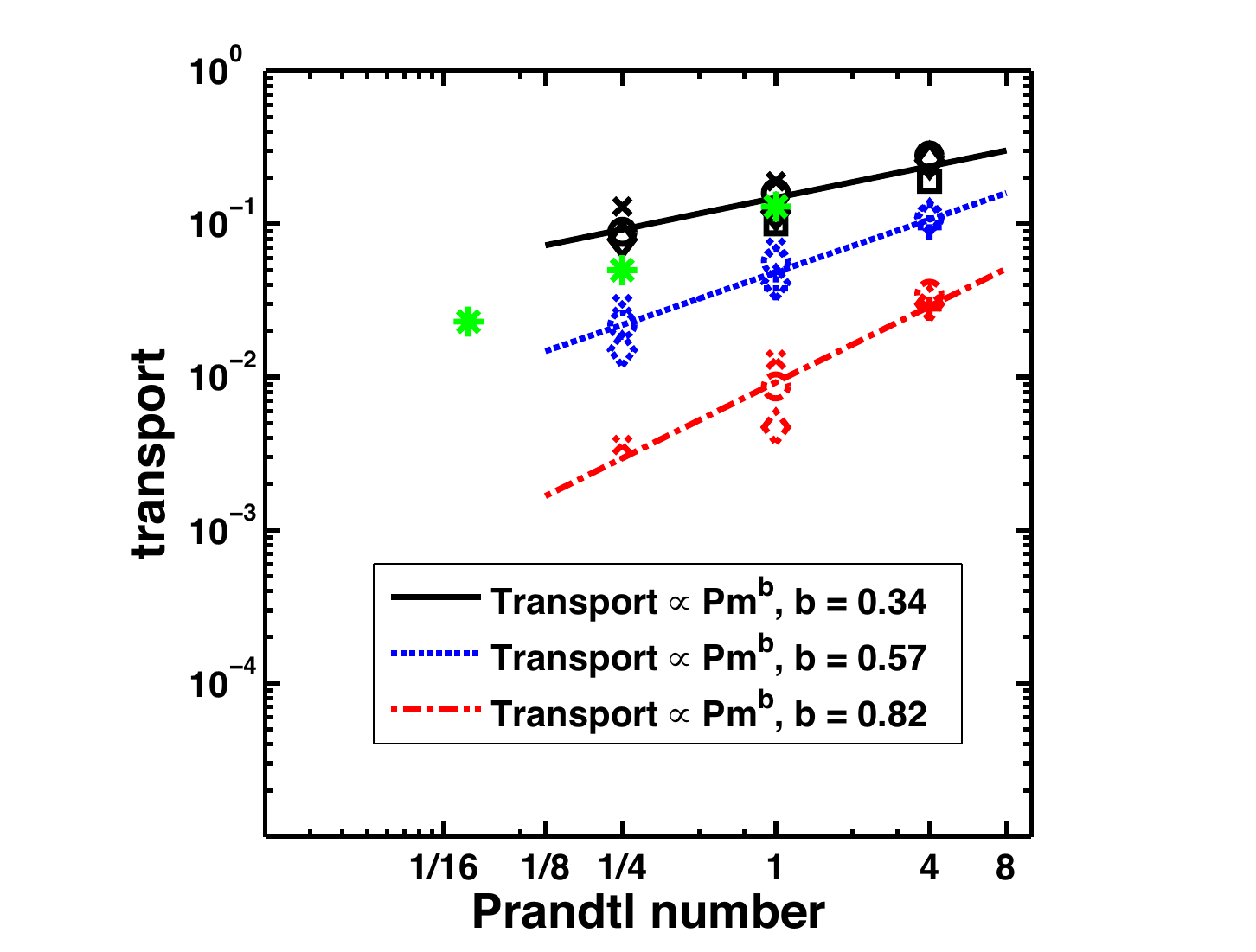}
\caption{Turbulent angular momentum transport coefficient $\alpha$
  vs.\ Prandtl number $Pm$ and plasma $\beta$ for various Reynolds
  numbers.  Solid, dotted and dot-dashed indicate $\beta=10^2$, $10^3$
  and $10^4$.  Squares, pluses, diamonds, circles and exes mark
  $Re=400$, $800$, $1600$, $3200$ and $6400$.  A power-law fit is
  shown for each value of $\beta$.  Stars mark 3~higher-resolution
  runs at $Re=20\,000$ and $\beta=10^3$
  \citep{longaretti&lesur10}.  \vspace*{-1.5mm}}
\label{fig:prandtl}
\end{figure}

To address the convergence issue, several investigations have included
a constant Ohmic resistivity, $\eta$, often along with a constant
kinematic viscosity $\nu$.  The result that has emerged is that both
dissipation coefficients are important: $\alpha$ is an increasing
function of their ratio, the magnetic Prandtl number $Pm=\nu/\eta$
(Fig.~\ref{fig:prandtl}).  This result is robust, having appeared in
homogeneous shearing boxes with net vertical
\citep{lesur&longaretti07, longaretti&lesur10} and toroidal
\citep{simon&hawley09} magnetic fluxes, with no net flux
\citep{fromangetal07} and in stratified boxes without net flux, though
it appears weaker in that case \citep{simonetal11}.  As an example, in
homogeneous boxes with a vertical magnetic field having $\beta=10^3$,
$\alpha$ ranges from $2\times 10^{-2}$ to $10^{-1}$ when $Pm$ varies
from $0.0625$ to $1$, all else being equal.  Recall that in the ideal
MHD limit (\S\ref{ideal_sat_sec}) this case shows numerical
convergence.  Clearly, numerical and physical convergence are two
different things and small-scale dissipation matters even in
situations that appear numerically converged, at least over the
limited range of Reynolds numbers that is numerically accessible.

These results raise questions about the flow structure in
protoplanetary disks' inner parts where the resistivity is much larger
than the viscosity \citep{2008ApJ...674..408B} ($Pm\ll 1)$.  The
problem is this regime is numerically challenging to investigate as it
requires extremely large spatial resolution: the smallest published
value of $Pm$ amounts to $1/16$.  Does $\alpha$ approach an asymptote
at lower $Pm$?  If an asymptote exists, how small is $\alpha$ there,
especially when stratification is included?  These questions' broad
implications make them a focus of ongoing research.

While these results highlight the properties of the MRI in
protoplanetary disks' inner parts, they are not directly applicable to
the planet-forming region of protoplanetary disks, where magnetic
coupling is good only in the surface layers.  In the scenario
suggested early on by \citet{1996apj...457..355g}, the disk is divided
into a laminar equatorial ``dead zone'' and MRI-turbulent surface
``active layers''.  Modeling this scenario, \citet{fleming&stone03}
used an Ohmic diffusivity that is a fixed function of time and height
above the equatorial plane.  The MRI is stabilized in the dead zone
while the active layers are turbulent.  The dead zone shows
hydrodynamical stresses excited by sound waves propagating from the
active layers, confirmed by \citet{2009ApJ...704.1239O} among others.
Another modification to the original scenario is that large-scale mean
magnetic fields can diffuse downward to transport angular momentum
within the dead zone itself \citep{2008ApJ...679L.131T}.  The $\alpha$
values reported are around $10^{-4}$.  Semi-analytic recipes are now
available describing the vertical structure of such layered disks
\citep{okuzumi&hirose11}.

Two further aspects of the layered disk picture have been investigated
intensively since the previous volume in this series.  The first
involves the changes over time in the Ohmic diffusivity profile as a
function the local flow properties \citep{2006a&a...445..223i,
  2007ApJ...659..729T, ilgner&nelson08}.  The second involves better
treating the disk thermodynamics \citep{2010MNRAS.409.1297F,
  hirose&turner11}.  The vertical distribution of heating differs
significantly from that assumed in $\alpha$-models.

\subsection{\textbf MRI With Hall Effect \label{hall_sat_sec}}

The first investigation of the MRI with Hall effect was performed by
\cite{2002apj...570..314s, 2002apj...577..534s} who found that MRI
turbulence with both Hall and Ohmic terms was rather similar to MRI
turbulence with the Ohmic term alone.  The stress depended weakly on
the Hall parameter, $\alpha$ increasing with $\eta_H$.  However,
\cite{2012MNRAS.422.2737W} pointed out that there was room for
stronger effects in the Hall-dominated regime where the Hall term is
the largest in the induction equation.

The problem was revisited by \cite{2013MNRAS.434.2295K} who showed
that under a Hall term 10-100$\times$ stronger than in
\citealt{2002apj...577..534s}, the MRI develops into self-sustaining
large-scale zonal magnetic field structures which produce only very
weak angular momentum transport ($\alpha\sim 10^{-5}$).  The
transition to this regime happens where $\Lambda_H\la5\,v_A/c_s$.
Therefore, even if the linear analysis shows fast growth at large
scales, the instability does not necessarily lead to MHD turbulence
and enhanced angular momentum transport.  Note that these results are
from unstratified shearing boxes.  It is thus too soon to rule out
Hall-dominated MRI as a source of turbulent transport in disks, since
stratification is likely to modify these results.

\subsection{\textbf MRI With Ambipolar Diffusion
  \label{ambipolar_sat_sec}}

Numerical implementation of ambipolar diffusion comes in two flavors.
In the two-fluid approach, ions and neutrals are treated as separate
fluids coupled by collisional drag, but typically not by chemistry,
with the magnetic field carried by the ions.  In the single-fluid
approach the neutrals are the fluid, with the ions, whose density is
determined by equilibrium chemistry, providing dissipation through
ion-neutral drag.  Generally speaking, the two-fluid approach applies
to well-ionized systems with slow recombination, while the
single-fluid approach applies in the opposite (``strong-coupling'')
limit.  Chemical modeling indicates the strong-coupling limit is
almost always valid in protostellar disks \citep{2011ApJ...739...50B}.
Early on \citet{1995ApJ...442..726M} incorporated ambipolar diffusion
in their single-fluid simulations, but a systematic study was lacking.
\citet{1998ApJ...501..758H} studied the MRI with ambipolar diffusion
using the two-fluid approach, so the results do not apply directly to
protostellar disks.

\citet{2011apj...736..144b} made a series of unstratified shearing-box
simulations of the MRI with ambipolar diffusion in the strong-coupling
limit.  They systematically explored the ambipolar diffusivity,
magnetic field strength and field geometry, finding consistent with
linear analysis (\S\ref{sec:linearMRIwAD}) that the MRI operates at
any value of $Am$ as long as the field is weak enough that the most
unstable mode fits within the disk scale height.  The saturation level
of the turbulence depends on field strength and configuration, and
when ambipolar diffusion is strong ($Am<10$), fields with comparable
net vertical and toroidal components are best, while a pure toroidal
field suppresses the MRI.  Summarizing their results, the saturation
level of the MRI turbulence with the most favorable field geometry is
\begin{equation}
  \alpha_\mathrm{max}\approx\frac{1}{2}\bigg[
    \bigg(\frac{50}{{Am}^{1.2}} \bigg)^2 +
    \bigg(\frac{8}{{Am}^{0.3}}+1\bigg)^2 \bigg]^{-1/2}.
\end{equation}

The next step has been to treat ambipolar diffusion in stratified
simulations.  Ambipolar diffusion is the dominant non-ideal MHD effect
in low-density regions, including the surface layer of the inner disk
together with the entire outer disk (\S\ref{sec:linearmri}).  For the
inner disk at $1-10$~AU, \citet{bs13} and \citet{2013ApJ...772...96B}
performed stratified shearing-box simulations that include both Ohmic
resistivity (dominant in the midplane) and ambipolar diffusion
(dominant at the disk surface) with diffusion coefficients based on
the equilibrium chemistry calculations of \citet{2011ApJ...739...50B},
as well as the FUV ionization model of \citet{2011ApJ...735....8P}.
They found that without net vertical magnetic flux, the system evolves
into extremely weak turbulence in the FUV layer, far too weak to drive
appreciable accretion.  When a net vertical magnetic flux is included,
even if the initial field configuration is subject to the MRI, the
instability is suppressed and the disk's whole vertical extent relaxes
into a laminar state, while a wind is launched and efficiently
extracts angular momentum from the disk (\S\ref{sec:winds}).  In both
cases the MRI is suppressed by ambipolar diffusion.  For the outer
disk at $\gtrsim30$~AU, \citet{simonetal13, sba13} performed similar
calculations and found first that net vertical magnetic flux is
essential otherwise the resulting MRI turbulence is too weak.  In the
presence of net vertical flux, the turbulence is stronger and takes
place mainly in the surface FUV layer.  Both results qualitatively
agree with expectations from unstratified simulations
\citep{2011apj...736..144b}, except that with increasing net vertical
magnetic flux, large-scale field comes to dominate over turbulent
field as the driver of angular momentum transport.

\subsection{\textbf Global Calculations
  \label{global_sat_sec}}

Local simulations are invaluable for studying turbulence, but their
reach is limited.  Thus several teams have embraced the challenges of
global simulations.  Resolutions do not yet approach those achieved in
the shearing box, where $>$100~cells per scale height is commonplace.
Consequently it is difficult at present to include small-scale
dissipation in global work.  The only successful example is
\citet{2010A&A...515A..70D}, where the Ohmic diffusion is treated.
Given all the subtleties uncovered when the non-ideal effects were
taken into account in shearing boxes, doubts remain about our ability
to characterize the saturation of the turbulence through global
calculations.  The global results to date should therefore be
interpreted with care.  Nevertheless, progress has been made thanks to
growing computer power and there have been claims that the
best-resolved models are numerically converged \citep{sorathiaetal12}.
Of course the lesson of the shearing box is that numerically-converged
simulations do not necessarily yield the correct stresses, as these
can depend on effects not treated \citep{2013ApJ...772..102H}.

The basic result of global simulations is boring and reassuring at the
same time: the transport is consistent with shearing box simulations
under similar conditions \citep{fromang&nelson06}.  The turbulent
velocity dispersion increases with height \citep{fromang&nelson09,
  flocketal11} as in shearing boxes \citep{2000apj...534..398m}.
There is growing evidence that large-scale flow features are similar
in wide shearing boxes \citep{simonetal12} and global calculations
\citep{2011MNRAS.416..361B, flocketal12}.  Likewise, the zonal flows,
axisymmetric variations in the orbital speed that persist for many
orbits, seem to be a natural outcome of MHD turbulence in both setups
\citep{johansenetal09, uribeetal11, 2013ApJ...763..117D}.

Some issues can only be addressed using global calculations.  For
example \citet{sorathiaetal10} reported that small patches of disk
behave as if threaded by a vertical net field, even if the total
vertical flux threading the disk as a whole vanishes.  Such fields
connect the disk atmosphere with its equatorial plane, suggesting that
coronal magnetic fields are central in determining the accretion flow
\citep{2011MNRAS.416..361B}.  Protoplanetary disks' large scale flow,
important for our understanding of particle transport, remains poorly
known.  The prediction from traditional $\alpha$-models of a
meridional circulation, with matter flowing outward near the midplane
while accretion proceeds through the upper layers, fails in fully
turbulent (and thus unrealistic) disks \citep{fromangetal11,
  flocketal11}.  It is likely that mixing within disks made up of both
active and dead zones differs from the models published so far.

Besides the resolution issue already mentioned, global simulations
suffer from two limitations of special importance.  First, the
equation of state is typically locally isothermal with no explicit
heating and cooling.  These are potentially important for the flow in
the disk corona, the structure of the disk interior, and the dead zone
nearest the star.  Second, no published global simulation of turbulent
protoplanetary disks yet includes a net vertical magnetic flux.  Given
the differences between shearing-box results with and without vertical
fluxes (\S\S\ref{ambipolar_sat_sec} and \ref{sec:winds}), such
simulations would be extremely valuable and should be a focus in
coming years.

\subsection{\textbf Outstanding Issues}

While small-scale dissipation has been (and still is!) studied in
detail, the results have mainly taught us to be humble: more than
20~years after the discovery of the MRI, we still do not know how the
turbulence saturates as a function of parameters such as the surface
density, temperature and magnetic flux.

The question of how disk dynamos work is still debated.  The results
of the last few years suggest that brute force is of limited interest
due to the enormous resolutions required.  An interesting alternative
is provided by the discovery of limit cycles near marginal stability
\citep{HR11}, which might bear some significance for fully developed
turbulence.

Perhaps most importantly for the transport, all of the results based
on the MRI linear properties agree in their predictions that
protoplanetary disks are MRI-inefficient at distances from the star
between about 1 and 10~AU \citep{2009apj...703.2152t,
  2011ApJ...739...50B, 2011ApJ...739...51B, 2011ApJ...735....8P,
  2011ApJ...727....2P, mohantyetal13, 2013ApJ...765..114D}.  In this
section, we have seen that many predictions of the linear stability
analyses carry over to the non-linear regime, with the notable
exception of the Hall effect.  Together these results suggest that
something is missing in our understanding of the basic processes
driving the accretion flows in the planet-forming region.  This has
revived interest in an old alternative transport mechanism, namely
disk winds, the focus of the next section.

%%%%%%%%%%%%%%%%%%%%%%%%%%%%%%%%%%%%%%%%%%%%%%%%%%%%%%%%%%%%%%%%%%%%%%%%
\section{\textbf{Disk-Driven Winds and MHD Turbulence}
  \label{sec:winds}}
In previous sections we explored the possibility of driving accretion
by transporting angular momentum inside the disk through the MRI.
However, this is not the only way accretion can occur.  An alternative
is to extract the angular momentum vertically by applying a torque at
the disk surface.  Such a torque usually comes from a large-scale wind
driven by magnetic processes.  This section is dedicated to these
winds' launching and their impact on accretion in protoplanetary
disks.  Jets formed as disk winds propagate to large distances are not
discussed in this section.  The interested reader may refer to the
chapter by Cabrit et al.

To understand how a cold wind can drive accretion, consider angular
momentum conservation in a thin disk
\begin{eqnarray}
\nonumber
\overline{\rho v_R}\frac{\partial \Omega R^2}{\partial
  R}&=&\frac{1}{R}\frac{\partial}{\partial R}R^2[\overline{B_\phi
    B_R}-\overline{\rho v_\phi v_R}]\\
\label{eq:angmom}
&& +R\Big[B_\phi B_z-\rho v_\phi v_z\Big]_{z=-h}^{+h},
\end{eqnarray}
where $\Omega$ is the Keplerian frequency, $h$ is the disk
scale-height and overbars mean $\int_{-h}^h\,\mathrm{d}z$.
Eq.~\ref{eq:angmom} shows that accretion (negative $\overline{\rho
  v_R}$) can be driven either by angular momentum transport inside the
disk (first term on the right-hand side) or by a torque exerted at the
disk surface (second term on the right-hand side).  It is the second
term which connects winds to accretion in the bulk of the disk.
Eq.~\ref{eq:angmom} also shows that a magnetically driven wind can
easily dominate accretion in thin disks.  Neglecting the kinetic term,
one easily deduces that the turbulent torque $\sim h B_\phi B_R$ while
the wind torque $\sim R B_\phi B_z$.  Assuming $B_\phi B_R\sim B_\phi
B_z$ within the disk, one finds that the torque due to the wind is
$\sim R/h\times$ larger than the torque due to turbulence.

\subsection{Launching Mechanism}
\label{sec:launch_mech}
Protoplanetary disks are thin ($h/R\lesssim 0.1$) and dynamically
cold, so winds cannot be launched exclusively by thermal processes at
the disk surface.  Driving a wind involves accelerating the flow to
escape speed, which requires some extra power source.  The power can
come either from heating the gas as it is ejected (see the chapter on
photoevaporation in this book) or from rotational energy extracted
using a large-scale magnetic field, a process known as
magnetocentrifugal acceleration \citep{BP82}.  The latter is
extensively described in the literature (see \citealt{s96} for an
introduction) so we here give only a brief qualitative picture.

Consider a frame co-rotating with the disk at radius $R_0$ from the
central star.  In this frame, fluid parcels feel an effective
potential combining the gravitational and centrifugal potentials.  If
poloidal $(R, z)$ magnetic field lines act as rigid wires for fluid
parcels, then a parcel initially at rest at $(R=R_0, z=0)$ can undergo
a runaway if the field line to which it is attached is more inclined
than a critical angle.  Along such a field line, the effective
potential \emph{decreases} with distance, leading to an acceleration
of magnetocentrifugal origin.  This yields the inclination angle
criterion $\theta>30^\circ$ for the disk-surface poloidal field with
respect to the vertical.  Here fluid parcels rotate at \emph{constant
  angular velocity} and so increase their specific angular momentum.

Magnetic field lines act as rigid wires so long as the poloidal flow
is slower than the local Alfv\'en speed --- that is, most of the
energy is contained in the magnetic fields.  The accelerating flow
eventually reaches a poloidal speed exceeding the Alfv\'en speed, at a
location known as the Alfv\'en point.  Above here, magnetocentrifugal
acceleration ceases and the flow winds up the field into toroidal
coils.  From this point on, fluid parcels rotate at \emph{constant
  specific angular momentum}.

Angular momentum is thus extracted from the disk first by the magnetic
torque at the disk surface ($B_\phi B_z$ in equation \ref{eq:angmom})
and stored in the magnetic field's toroidal component.  As
acceleration proceeds, the angular momentum is slowly transferred to
the ejected material until the Alfv\'en point is reached.  Winds
therefore efficiently drive accretion by purely and simply removing
the angular momentum from the disk.  Moreover, one can see easily that
the efficiency of this process is directly connected to the disk's
magnetic field strength, a stronger field leading to a bigger torque
and therefore faster accretion.

We have seen that a wind implies an accretion flow inside the disk.
The poloidal field lines threading the flow are accreted too, if ideal
MHD is valid.  Such a flow must soon stop once a large magnetic flux
accumulates near the disk's center.  A long-lived wind requires the
flow to slip past the field lines.  This is usually modeled assuming
the disk is either turbulent, producing a turbulent diffusivity
\citep{fp93,fp95,f97}, or subject to strong ambipolar diffusion
\citep{wk93,ksw10,skw11}, a reasonable assumption at large distances
($\gtrsim 10\,\mathrm{AU}$).

These very simple arguments show three basic requirements for a steady
magnetocentrifugal wind:
\begin{itemize}
\item The field is strong enough for the poloidal component to act as
  a rigid wire.  This means that at the base of the wind the plasma
  $\beta\lesssim 1$, where the magnetic pressure is that in the mean
  poloidal field.  Fluctuations due to turbulence can exceed the mean
  field, but are not directly relevant for the existence of a steady
  wind.
\item The field is sufficiently inclined, with $\theta>30^\circ$.
\item Accreting gas slips through magnetic field lines, via either
  turbulent diffusion, or Ohmic or ambipolar diffusion.
\end{itemize}

\subsection{Local Disk Wind Solutions}
A natural question is whether the wind can coexist with turbulence
inside the disk.  However, numerically spanning both the small
turbulent eddy scales and the large global scales associated with the
wind propagation is a challenge.  Two complementary approaches are
used: (1) Local methods such as the shearing box involve treating the
disk up to a few scale heights while neglecting the global geometry.
(2) The global approach means approximating turbulence by some crude
mean-field method.  We focus here on the first approach since our
topic is disk transport, but we also discuss the connection to global
solutions to demonstrate the limitations of local solutions.

The shearing box outlined in \S\ref{sec:saturation} is shear-periodic
in the radial or $x$-direction and periodic in the toroidal or
$y$-direction.  The vertical or $z$-boundaries have outflow conditions
to cope with winds.  This approach has several limitations:
\begin{description}
\item{Symmetries:} The shearing box does not specify whether the
  central object lies at $x\rightarrow +\infty$ or $x\rightarrow
  -\infty$, leaving the two equivalent.  This symmetry is apparent in
  shearing box results, which can exhibit winds bent toward either
  large or small~$x$.
\item{Magnetic flux transport:} Because curvature terms are neglected
  in the shearing box, magnetic flux tubes can cross the box again and
  again without the fields piling up anywhere, since the flux escaping
  matches that entering on the box's far side.  In other words,
  shearing boxes allow spurious solutions with constant $E_\phi\ne0$.
  This unrealistic situation should be kept in mind when one compares
  local and global wind solutions.
\item{Boundary conditions:} The third and most stringent limitation
  comes from the effective potential (gravitational plus centrifugal)
  in the shearing box being a local expansion of the real potential
  around $R_0$ of the form
  \begin{equation}
    \psi\propto -\frac{3}{2}x^2+\frac{1}{2}z^2,
  \end{equation}
  where $x$ and $z$ are respectively local equivalents of $R-R_0$ and
  $z$.  Clearly this potential does not allow a particle to escape to
  $z\rightarrow\infty$ if $x$ is kept finite.  Because
  magnetocentrifugal wind solutions cannot propagate radially past the
  Alfv\'en radius, disk winds \emph{cannot be gravitationally unbound}
  in a finite shearing box.  The winds that nevertheless appear owe
  their existence to the outflow vertical boundary conditions.
\end{description}
Below we show how these limitations affect the numerical results.

All the local numerical wind solutions obtained in recent years can be
understood in the framework of the magnetocentrifugal acceleration
mechanism described above.  Here we distinguish ``strong field''
solutions, for which magnetic and thermal pressure are near
equipartition $\bmid\sim 1$ in the midplane, from ``weak field''
solutions where thermal pressure dominates at the midplane $\bmid\gg
1$ (but not necessarily in the disk corona).  We furthermore separate
the solutions by the non-ideal effects included.

\paragraph{Ideal and weakly resistive winds}
\label{sec:idealMHD}
Ideal MHD solutions are the simplest obtainable, but of course neglect
the non-ideal effects which can be important in protoplanetary disks.
Moreover, under ideal MHD the numerical convergence of MRI turbulence
can be an issue (\S\ref{sec:saturation}).  To address these
limitations, some solutions have been computed in the resistive MHD
framework.  A weak resistivity and viscosity seem not to affect the
wind dynamics \citep{fllo13}.

The first ideal-MHD solutions exhibiting both a disk wind and the MRI
on weak fields ($\bmid\gg 1$) were obtained by \citet{si09}.  These
involve weak poloidal fields $\bmid=10^6$---$10^4$.  Standard MRI
turbulence appears and a short-lived magnetocentrifugal wind slowly
empties the domain since the shearing box boundaries allow no
replenishment by accretion flow.  Because the midplane fields are
weak, the wind is launched about two scale-heights up, so that
$\beta\sim 1$ at the launching surface as expected from the
phenomenology presented above.  The wind mass loss rate $\dot{M}_W$
increases steeply as $\beta$ decreases, almost emptying the box after
75~rotations for $\beta=10^4$.  \citet{smi10} explored the
consequences of such a wind for global disk models, demonstrating that
a wind could empty the inner disk and create an inner hole.
\citet{bs12} explored stronger fields up to $\beta=100$.  They
confirmed some of the \citet{si09} results, showing that
$\dot{M}_W\propto 1/\beta$ holds for the $\beta$ values they explored.
However, \citet{fllo13} and \citet{bs12} demonstrated that $\dot{M}_W$
was highly sensitive to the domain height: $\dot{M}_W\propto 1/L_z$
where $L_z$ is the box vertical size.  This is a clear signature of
the third shearing box limitation detailed above.  Box simulations
yield unreliable results for the mass flow rate extracted by disk
winds and the efficiency of disk dispersal.

The shearing box's extra symmetries also allow the wind to switch
erratically between ejection toward larger and smaller radii
\citep{fllo13, bs13}.  The configuration switchings happen
independently above and below the disk.  For example, the upper disk
can drive a wind toward larger radii while the lower disk drives a
wind toward smaller radii.  Such artificial outcomes are
unsatisfactory for a global overview of the wind structure.

The strong-field regime $\bmid\sim 1$ is well-studied for jet
formation because it is the only one where self-similar global wind
solutions exist \citep{f97}.  This regime has also been explored in
the local approximation.  \citet{o12} obtained 1-D quasi-steady
strong-field wind solutions for which $\dot{M}_W$ \emph{decreases}
sharply when $\beta$ decreases.  Wind solutions possibly even cease to
exist for sufficiently strong fields ($\bmid\simeq0.3$).  This study
was revisited in 3-D by \citet{lfo13} in a regime where wind solutions
of \citet{o12} coexist with large-scale MRI modes \citep{lfg10}.  They
showed that MRI modes naturally evolve into quasi-steady wind
solutions when the field is strong enough ($\beta\sim10$), making a
natural connection between the MRI and strong winds.  However,
similarly to the weak field case, $\dot{M}_W$ depends on the box
vertical extent, casting doubt on the conclusions.  \citet{m12} and
\citet{lfo13} showed that the quasi-steady wind solutions of
\citet{o12} were radially unstable, producing time-dependent wind
solutions.  Several mechanisms potentially lead to unstable winds
\citep{lpp94b,ls95,cs02}.  However none seems entirely compatible with
numerical results \citep{m12,lfo13}.

In all these numerical solutions, the toroidal electromotive force
$E_\phi\ne0$ and magnetic flux tubes accrete.  This is to be expected
since the solutions are computed in ideal or weakly resistive regimes.

\paragraph{Winds with ambipolar diffusion and Hall effect}
The Hall effect and ambipolar diffusion are believed to dominate in
much of protoplanetary disks (\S\ref{sec:linearmri}).  Ambipolar
diffusion is specially important since it lets the gas cross magnetic
field lines rather than sweeping them along.  In principle this avoids
the flux accumulation problem faced by the ideal-MHD and weakly
resistive models described in \S\ref{sec:idealMHD}.
 
Local solutions including both Hall effect and ambipolar diffusion
were initially presented by \citet{wk93} who treated the ions,
electrons and neutrals as independent fluids coupled together by a
drag force.  With this approach, the neutrals (which make up most of
the mass) can accrete without advecting field lines (which are
attached to the charged species).  \citet{wk93} demonstrated steady
wind solutions in the $\bmid\simeq 1$ regime, both with and without
accreting magnetic field lines.  Surprisingly, solutions with and
without magnetic accretion are similar, indicating the magnetically
accreting solution presented in \S\ref{sec:idealMHD} might be
qualitatively correct.

Comparable winds can be launched when the Hall and Ohmic diffusivities
are large \citep{ksw10,skw11}.  Interestingly, the mass-loss rate and
torque depend on the poloidal field \emph{polarity} (i.e.\ the sign of
$\bf{B\cdot \Omega}$).  In particular, some wind solutions are
obtained only for positive polarities.

The problem of outflows in the presence of ambipolar diffusion was
revisited by \citet{bs13}.  Using a complex chemical network to
compute ambipolar and Ohmic diffusivities, they found midplane
ambipolar diffusion several orders of magnitude stronger than
anticipated by \citet{wk93} and concluded that non-ideal effects
suppress MRI turbulence at 1~AU.  However, even poloidal fields as
weak as $\bmid=10^5$ trigger a quasi-steady wind starting about
4~scale-heights above the midplane, where ambipolar diffusion is
significantly reduced thanks to FUV ionization.  While the shearing
box is ambiguous regarding the direction to the star, global
magnetocentrifugal wind solutions generally have the horizontal
components of the magnetic field changing sign across the disk.  In
the local weak-field ambipolar-dominated calculations, the horizontal
field flips in a thin layer offset from the midplane by several
scale-heights.  This layer, a fraction of a scale-height thick,
receives the wind torque and carries the entire accretion flow.  Based
on these results, \citet{2013ApJ...772...96B} proposed that the
accretion rates in the planet-forming region at $\sim$1--10~AU can be
explained by such a steady wind alone, the bulk of the disk being
essentially quiet.  The ambipolar-diffusion-dominated outer disk is
likely MRI turbulent but can also launch an outflow \citep{sba13}.
 
\paragraph{Connecting local and global solutions}
In one of the first attempts at connecting local and global solutions,
\citet{wk93} matched their local results to a global \cite{BP82} wind,
ruling out some of the local solutions.  Similarly, \citet{lfo13}
compared their local solution against global solutions computed
\emph{\`a la} \citet{fp95}.  There was qualitative agreement but
significant differences remained, e.g. in mass loss rate and $\bmid$.
Interestingly they found no global solutions for fields as weak as
they considered in the local solutions, suggesting that if a global
wind exists for given parameters, then the local approach will catch
the right solution.  However, the local approach might also catch
solutions which simply do not exist in global geometry.  These
spurious winds are impossible to distinguish from ``real'' winds
solely by a local approach.  Future work on this topic will therefore
have to extend the recent findings on local solutions to global
geometries.
 
\subsection{Global Disk Wind Solutions}
To overcome the limitations of local solutions, one can perform global
calculations.  Here however a scale separation problem occurs.  On one
hand, any turbulence must be resolved.  On the other hand, the
solution must be computed far from the disk in the vertical direction
(typically to 10s or even 100s of disk radii).  For this reason, all
global disk wind solutions have been obtained in a 2.5-D approximation
with turbulent motions either ignored or treated as viscosities and
resistivities.  This approximation is brought into question by 3-D
results on torus accretion by a black hole \citep{bhk09}.  The main
difficulties with the global models are two-fold:
\begin{itemize}
\item They need a strong magnetic flux in the disk midplane
  ($\bmid\sim1$), which can either build up as flux accretes from the
  outer disk (a scenario with several problems, see \citealt{lpp94})
  or be put in place when the disk forms (in which case the entire
  question of jet formation is linked to the disk formation scenario).
\item They rely on strong resistivities.  A resistivity $\eta_t$,
  supposed to mimic the local effects of 3-D turbulence, is quantified
  by the parameter $\alpha_{m}$ in $\eta_t=\alpha_{m}c_sH$.  To avoid
  magnetic flux accumulating, these models require $\alpha_{m}\sim 1$.
  However no physical mechanism justifies such a high turbulent
  diffusivity with $\bmid\sim 1$.  In particular, MRI turbulence is
  believed to produce at best $\alpha_{m}\sim 0.1$
  \citep{2009A&A...504..309L}.
\end{itemize}
The behaviour of global models as a function of the turbulent
resistivity was investigated by \citet{zfr07} in 2.5-D simulations.
For $\alpha_{m}=1$ and $\bmid=1$, they found quasi-steady ejection
resembling the self-similar models of \citet{f97}.  Lower values of
$\alpha_{m}$ invariably lead to unsteady ejection, probably due to
episodes when magnetic flux accretes followed by sudden releases of
the flux in numerical reconnection or ejection events.  This indicates
that ejection is possible for weak diffusivities, even though the
resulting winds and jets are likely to be highly variable.

Issues of disk magnetisation were studied by \citet{tfm09} and
\citet{mfz10}.  They demonstrated the possibility of launching
unsteady magnetically-driven winds up to $\bmid\sim 500$.  However
these models rely on large diffusivities with $\alpha_{m}\simeq 1$
which are probably too high to be realistic if turbulence is driven
solely by the MRI.  They also pointed out the sensitivity of the mass
loading rate (fraction of the mass which is ejected in the wind) to
the resolution.  The high mass loading rate achieved in these
simulations is likely due to numerical diffusion of gas at the top of
the disk, where the launching occurs.

All the models considered above are ``cold'', with the turbulent
heating canceled by radiative cooling.  \citet{tfm13} showed
explicitly that if weak turbulent heating is not balanced by radiative
cooling at the disk surface, the mass loading rate and ejection
efficiency are significantly affected.  This demonstrates the
sensitivity of winds to thermodynamics near the launch point, which
must be modeled accurately if we are to predict winds' large-scale
properties and match jet observations (see chapter on jets by Frank et
al.\ in this book).

\subsection{Outstanding Issues}
The new disk wind scenario differs significantly from the $\bmid\simeq
1$ picture investigated in the 1990s.  The magnetization is weak, with
$\bmid\sim 10^4$, and the ambipolar diffusivity varies with height.
This scenario is still developing and global modeling may soon clarify
the picture.  Outflowing material could shield the disk from ionizing
stellar photons \citep{2003MNRAS.342..427F, 2012A&A...538A...2P},
preventing wind launching.  The winds' sensitivity to radiative
transfer and chemistry must be investigated further.  Finally we note
that this scenario is unlikely to explain the protostellar jets
launched within 1~AU, where non-ideal MHD effects are weak thanks to
thermal ionization.

%%%%%%%%%%%%%%%%%%%%%%%%%%%%%%%%%%%%%%%%%%%%%%%%%%%%%%%%%%%%%%%%%%%%%%%%
\section{\textbf{Self-Gravitating Turbulence}
  \label{sec:GI}}
Self-gravitating, non-axisymmetric structures transport angular
momentum with a flux density
\begin{equation}
W_{r\phi} = \int dz \left( \frac{g_R g_\phi}{4\pi G} + \rho v_R \delta
v_\phi \right)
\end{equation}
where $g_R$ and $g_\phi$ are the radial and azimuthal components of
the gravitational acceleration.  This transport is {\em non-local} in
the sense that it can couple together parts of the disk that are
widely separated in radius \citep{BalbPap99}.  Gravitational torques
may, however, be described by an effective $\alpha$, if the
characteristic scale $\lambda_{char}$ is small compared to the local
radius \citep{LodatoRice04, LodatoRice05, VoroBasu09, cossins09,
  Forgan11, Michael12} or equivalently the disk mass is small compared
to the star \citep{Voro10b}.

Self-gravitating disk structures are detectable in the millimeter
continuum and lines in nearby ($d \lesssim 150$~pc) systems with ALMA
\citep[e.g.][]{cossins10,douglas2013}, although self-gravitating disks
are likely to be found in young, short-lived systems, and are
therefore expected to be rare.  If they are found, torques and a disk
evolution timescale can be measured directly from a surface density
map, e.g.\ \citet{gnedin1995}, and molecular line observations may
allow us to measure the ratio of vertical to in-plane velocity
dispersion well enough to separate the values expected for
gravitationally-driven and MRI-driven turbulence
\citep{2012MNRAS.426.2419F}.

When is the disk self-gravitating?  Self-gravity becomes important
when the Toomre parameter
\begin{equation}
Q \equiv \frac{c_s\kappa}{\pi G \Sigma} \sim 1
\end{equation}
where $c_s$ is the sound speed, $\kappa$ the epicyclic frequency which
$\simeq \Omega$ in a Keplerian disk, and $\Sigma$ the surface density.
If we define the scale height $H \simeq c_s/\Omega$ and the disk mass
$M_d \equiv \pi R^2\Sigma$, then we can rewrite $Q$ to obtain the
useful relation $Q = (R/H) M_d/M_*$.

The $Q$ parameter describes the balance between the stabilizing
influence of pressure ($c_s$) and rotation ($\kappa$) and the
destablizing influence of self-gravity ($G\Sigma$).  Pressure and
rotation are weakest in comparison to self-gravity at a characteristic
wavelength
\begin{equation}
\lambda_c \equiv \frac{2 c_s}{G \Sigma} = 2 \pi Q H.
\end{equation}
Measuring this characteristic scale in a face-on disk may then be a
direct probe of the surface density, through $Q$; numerical
experiments show that even in the non-linear regime much of the power
is concentrated near $\lambda_c$ \citep{cossins09}.

The formal analysis leading to the conclusion that $Q>1$ implies
stability has formal limitations: it relies on an {\em axisymmetric}
analysis of a zero-thickness disk, and requires that $\lambda_c \ll r$
(it is a {\em local}, or lowest order {\em WKB}, analysis).  But a
self-gravitating disk is most responsive to non-axisymmetric
perturbations; finite thickness is stabilizing \citep{GoldBell65,
  MamRice10}; and long-wavelength modes can be stabilized or
destabilized by global features in the disk \citep[e.g.][]{Mesch08}.
In the end, a good rule of thumb is that $Q \lesssim 2$ disks are
strongly self-gravitating.

The outcome of gravitational instability depends on how the disk is
driven unstable.  In one commonly analyzed model the disk is driven
unstable by cooling that is characterized entirely by the timescale
$\tau_{cool}$ \citep{Gam01}, or equivalently by the dimensionless
$\beta_c = \tau_{cool}\Omega$.  Then there is a $\beta_\mathrm{crit}$
\citep{ShlosBeg87} dividing models that fragment immediately ($\beta_c
< \beta_\mathrm{crit} \simeq 3$) from those that enter a quasi-steady, or
gravito-turbulent, state at $\beta_c > \beta_\mathrm{crit}$.  Since
cooling times typically decrease outward in protoplanetary disks, this
implies that there is a critical radius at a few tens of AU beyond
which a massive disk can fragment \citep{matzlev05,rafikov05}.  The
disk can also be driven unstable by mass loading via infall.  In this
case whether the disk fragments depends on the ratio of infall to
accretion timescales
\citep[e.g.][]{Boley09,Rafikov09,Kratter10,Harsono12}.

In a gravito-turbulent state the disk may settle into a condition of
marginal stability, with $Q$ slightly greater than the marginally
stable value \citep{Pac78,2007MNRAS.381.1009V}.  Transport in the
gravito-turbulent state can be described by an effective $\alpha$, and
is predominantly local, if $\lambda_c/R \lesssim 1$.  If angular
momentum transport can be described by an effective $\alpha$, its
value is $(4/9)\beta_c^{-1}/(\gamma(\gamma-1))$ \citep{Gam01}.  Then a
minimum $\beta_c$ implies a maximum $\alpha$ in the gravito-turbulent
state; at larger values the disk fragments.  The
$\alpha(\tau_{cool},\Omega)$ relationship also applies to more
complicated cooling laws, but if there are order-unity variations in
the surface density then it may be difficult to estimate
$\tau_{cool}$.

Angular momentum transport will not cease if the disk fragments. Even
low-mass bound objects will generate wakes in a self-gravitating disk
due to tidal interactions \citep{JulToom66} and superposition of these
wakes transports angular momentum \citep{JohnGood06}.  Massive,
embedded objects can lead to rapid disk evolution
\citep[e.g.][]{Krum07, Voro10b}, or a complicated interplay between
disk and planetary orbital evolution \citep{Boley09, Michael11}.

The critical value of $\beta_c$ remains unclear \citep{MeruBate11}.
Numerical outcomes are sensitive to numerical diffusion
\citep{Pard11,Pard12,MeruBate12}, the use of 2-D vs.\ 3-D models,
gravitational softening \citep{Muller12}, cooling with realistic
opacities \citep[e.g.][]{JohnGam03}, and perhaps most important,
irradiation \citep[e.g.][]{matzlev05, Clarke07, Cai08, Rice11,
  Forgan13}, which at the least increases the temperature and
therefore stabilizes the disk.

\begin{figure}[tb!]
   \centering
   \includegraphics[width=0.9\linewidth]{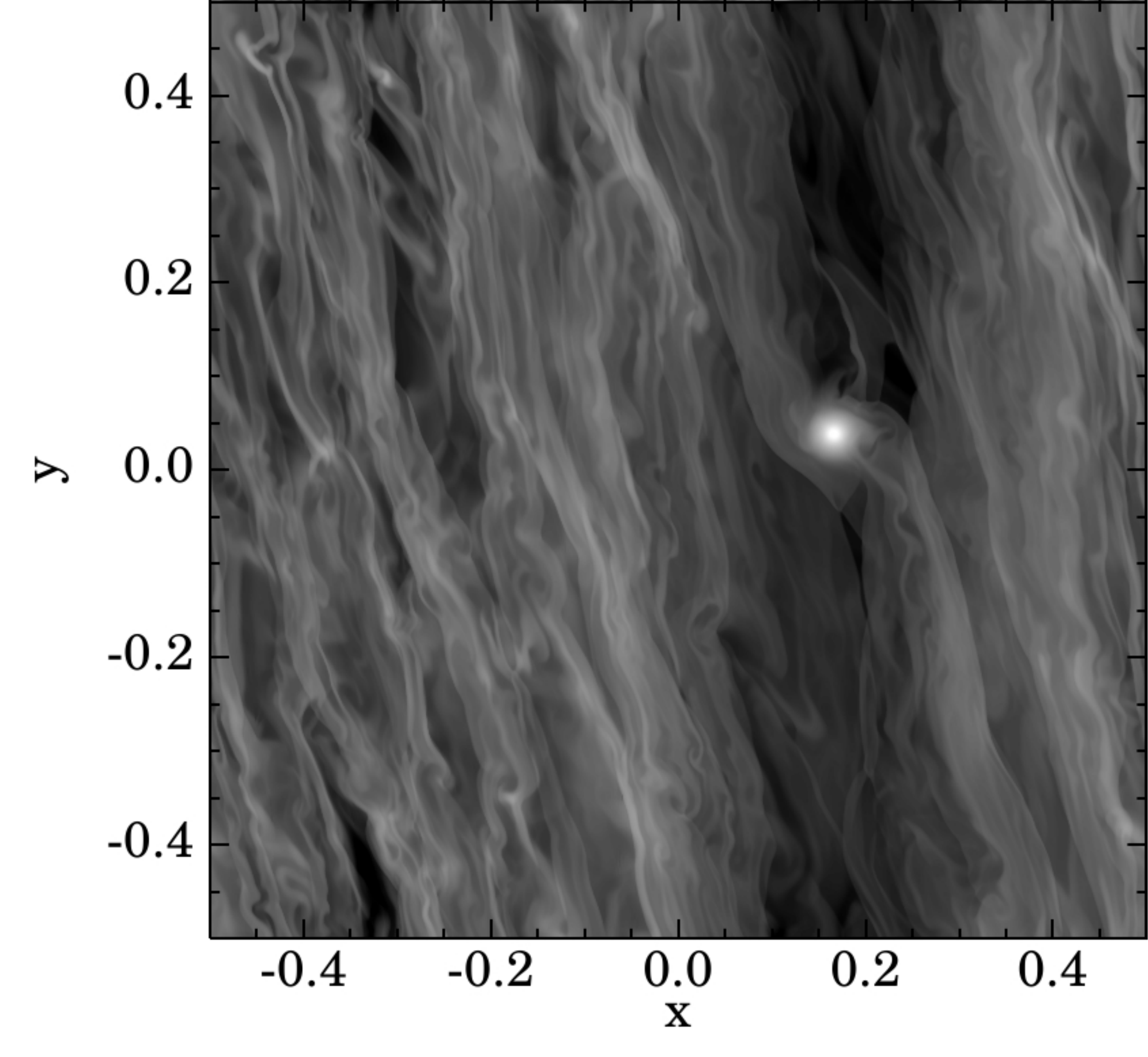}
   \caption{A 2-D (razor-thin) numerical model of a cooling,
     gravitationally unstable disk with $\beta_c = 5$ is in an
     apparently steady gravito-turbulent state at $\Omega t = 100$ but
     by $\Omega t = 400$ produces a bound object as shown.  Shading
     indicates surface density in units of the mean.  From
     \citet{Pard12}.  \vspace*{-1.5mm}  }
   \label{fig:paard}
\end{figure}

The most surprising new discovery about fragmentation is the finding
by \citet{Pard12} \citep[see also][]{HopChris13}, stimulated by work
of \citet{MeruBate11}, that a 2-D local disk model can still fragment
for $\beta_c \gg \beta_\mathrm{crit}$, if one waits long enough!  This
is illustrated in Fig.~\ref{fig:paard}, showing an apparently steady
gravito-turbulent state later fragments at high enough resolution.
The result is best described as stochastic fragmentation: for $\beta_c
> \beta_\mathrm{crit}$ the disk has a finite probability of producing
a bound object per unit area per unit time.  It is not yet known
whether this extends to models in 3-D, or with more realistic cooling.
Stochastic fragmentation may revive direct planet formation by
gravitational instability for disks with $\beta_c >
\beta_\mathrm{crit}$ \citep[e.g.][]{Boss1997}, or it may imply a
$\beta_\mathrm{crit,eff} > \beta_\mathrm{crit}$ where the
fragmentation time is the disk lifetime.

Evidently there are many open questions about angular momentum transport
by gravitational instability.  We think two in particular merit further
exploration: (1) magnetized, self-gravitating disks.  Does magnetically
driven turbulence (in surface layers, or throughout the disk) enhance or
limit the development of bound structures in protoplanetary disks?
Pioneering simulations \citep{Fromang2005} need to be revisited at
higher numerical resolution; (2) thermodynamically self-consistent disks
with realistic opacities and external irradiation.  While much work has
already been done \citep{Boss07,Cai08,SteimanCam13}
tracking cooling inside proto-fragments requires very high spatial
resolution.

%%%%%%%%%%%%%%%%%%%%%%%%%%%%%%%%%%%%%%%%%%%%%%%%%%%%%%%%%%%%%%%%%%%%%%%%
\section{\textbf{Hydrodynamical Processes or Disk Weather}
  \label{sec:weather}}
Our understanding of the flows triggered solely by gas pressure and
viscous forces has advanced greatly since the previous book in this
series.  Giant vortices, which may accelerate the planet formation
process \citep{BargeSommeria95, Tanga96, KlahrBodenheimer06}, have
gained observational support through ALMA observations of emission
perhaps from mm-sized particles concentrated to one side of a young
star \citep{2013Sci...340.1199V}.  The feature resembles predictions
of the vortices' appearance by \citet{WolfKLahr02} but has higher
contrast since the big particles concentrate more strongly than the
sub-$\mu$m grains in the model \citep{2013ApJ...775...17L}.

Hydrodynamical turbulence is less robust than the MRI and self-gravity
--- otherwise it would have appeared in numerical models years ago.
When magnetic fields couple to the gas, the MRI governs the flow
\citep{lk11, 2013MNRAS.435.2610N}.  Thus the most promising sites for
non-magnetic instabilities are the magnetically-decoupled dead zones.
It seems likely that the Solar nebula at 1--10~AU from the young Sun
was a suitable location.

Since protostellar disks have enormous Reynolds numbers $Re>10^9$, it
might seem that their orbital shear would easily lapse into
turbulence.  However as rotating fluids, the disks are a special case
of the Taylor-Couette flow, which is linearly stable when the angular
momentum increases with radius, as it does on the Keplerian rotation
curve.  This is embodied in the Rayleigh stability criterion
$\partial_R \left(\Omega^2 R^4\right) > 0$ \citep{Rayleigh17} at
radius $R$ and rotation frequency $\Omega$.  Based on contradictory
Taylor-Couette experiments spanning almost a century \citep{Taylor23,
  Wendt33, Taylor36, Schultz-Grunow59, Lathrop92, Richard99, Ji06,
  Paoletti12, Schartman12}, disks could be non-linearly either stable
or unstable.  Yet the experiments depart significantly from
astrophysical disks, especially at the vertical boundaries.  The
containers' upper and lower lids are argued to drive any turbulence
observed in the experiments \citep{Avila12}.  Also, even if accretion
disks become turbulent at some critical Reynolds number above the
maximum $Re\sim 2\times 10^4$ achieved in numerical simulations, the
resulting turbulent viscosity must still be relatively low, in fact
inversely proportional to the critical Reynolds number \citep{ll05}.
Thus we are compelled to investigate instabilities that occur under
flow conditions relevant to protostellar disks and can be studied
numerically.

Thermal convection perpendicular to the disk's midplane was once a
favorite to cause turbulence.  Yet it suffers a self-consistency
problem, producing too little heating to sustain the vertical entropy
gradient that drives it \citep[as reviewed by][]{Klahr07}.  Convection
pumped artificially by heating the midplane can however transport
angular momentum outward \citep{kb03, lo10}.

More promising are three classes of instabilities driven by radial
stratification.  We consider in turn the gradients in pressure,
temperature and entropy.  (1) Radial pressure gradients shift the
rotation profile away from Keplerian.  (2) Radial temperature
gradients with Coriolis forces yield vertical shear, or orbital speed
gradients in height, called ``thermal winds'' in geophysical settings.
(3) Radial entropy gradients make the disk gas radially buoyant, but
not so buoyant as to directly overcome the angular momentum barrier.
In the next three sections we briefly describe what is known about the
instabilities arising in these settings.

\subsection{Rossby-Wave Instability}
A bump in the radial pressure profile yields a local extremum in the
rotation speed, or more precisely in the vortensity (vorticity divided
by surface density).  The extremum is linearly unstable to Rossby-wave
instability (RWI), which develops into vortices in a few orbits
\citep{Colgate99, Li00, Li01}.  Since the orbital speed's radial
gradient is key, the RWI is a special kind of Kelvin-Helmholtz
instability found in rotating systems.  While the RWI develops under
conditions readily studied in numerical simulations \citep{Meheut10,
  Meheut12a, Meheut12b} the unstable initial state has to be explained
in the first place.  Suitable conditions can occur near the outer edge
of a gap opened in the disk by a planet \citep{Balmforth01,
  deVal-Borro07}.  In another scenario, a pressure bump forms at a
dead zone's inner edge \citep{Varniere06, 2009A&A...497..869L}.
Varni{\`e}re used a radially-varying $\alpha$ value for this
experiment, while \citet{LyraMacMlow12} performed a global non-ideal
MHD simulation with radially increasing resistivity and also found a
pressure bump, and subsequently a large vortex.  In conclusion, the
RWI does not make vortices in an unperturbed laminar disk, but
requires either planets or neighboring MHD active zones.

\subsection{Goldreich-Schubert-Fricke Instability}
In protostellar disks the temperature and density vary independently.
Temperatures are set by starlight heating outside a central
accretion-dominated region \citep{2007prpl.conf..555D}, while
densities are determined by the accretion flow.  Surfaces of constant
pressure therefore tilt relative to surfaces of constant density,
making protostellar disks at least in part baroclinic.  The baroclinic
regions reach a balance between the gas pressure, gravity and
rotational forces by forming a ``thermal wind'' in which the orbital
frequency increases with the distance from the midplane
\citep{Tassoul07, fromangetal11}.  The increase is quadratic where the
temperature is constant with height.  The resulting shear is locally
stable against the Kelvin-Helmholtz instability (KHI) owing to the
vertical density stratification \citep{2002A&A...391..781R}.  KHI can
nevertheless occur if the embedded solid particles sediment with
respect to the gas, forming a thin midplane layer rotating at
Keplerian speed.  The shear relative to the slower-rotating gas can
yield Richardson numbers low enough for KHI to produce mild local
turbulence \citep{1993prpl.conf.1031W, 2006ApJ...643.1219J,
  2009ApJ...691..907B, 2010ApJ...718.1367L}.

Accretion disks' radial shear also permits an instability first
studied for rotating stars' interiors, the Goldreich-Schubert-Fricke
(GSF) instability \citep{Goldreich67, Fricke68}.  This can be
triggered when the radial shear is Rayleigh stable but receives a
boost from vertical shear.  Under the combined shear, there exist
paths inclined slightly with respect to the vertical, along which the
specific angular momentum is constant.  Exchanging gas parcels along
these paths can be unstable if the disk's vertical stratification is
neutrally buoyant, with Brunt-V\"ais\"al\"a frequency zero.  Any real,
non-zero buoyancy frequency stabilizes the exchange of gas parcels.
The instability thus occurs only if the disk either is adiabatically
stratified, or relaxes thermally on very short time scales.
Quick-enough heating and cooling can overcome the stable vertical
density stratification.  Indeed in the two limiting cases of
instantaneous cooling \citep{1998MNRAS.294..399U} and isentropic or
uniform-entropy stratification \citep{2002A&A...391..781R}, the
buoyancy frequency and thus the Richardson number are effectively zero
and the Rayleigh stability criterion generalizes to
$\partial_R\Omega^2(R,z)R^4 - (k_R/k_z) R^4 \partial_z \Omega^2(R,z) >
0$ \citep{2003A&A...404..397U, 2013MNRAS.435.2610N}.  Hence perturbing
the vertical rotation profile can lead to weak, transient turbulence
until the shear decays, even in an isothermal flow with an unsheared
background state \citep{2004A&A...426..755A}.  When the vertical shear
is part of the background profile, the disk undergoes narrow
near-vertical overturning motions, leading to sustained accretion
stresses at levels $\alpha\sim 10^{-3}$ \citep{2013MNRAS.435.2610N}.
The instability is suppressed when thermal relaxation enforces an
isothermal atmosphere slower than $0.01\Omega^{-1}$.  The threshold is
more easily reached when the vertical structure is driven towards
isentropy: instability occurs for thermal relaxation times up to
$1\Omega^{-1}$.  Candidates for regions with thermodynamics suitable
for the GSF are disks with moderate optical depths, and in particular
the outer reaches of T~Tauri disks well above the midplane.

\subsection{Baroclinic Vortex Formation}

Vortices can also arise in disks having a radial entropy gradient
\citep{kb03}.  Unlike the baroclinic instability of stars and
planetary atmospheres, which relies on the vertical shear and seems to
be suppressed in disks by the radial shear, the baroclinic instability
in disks depends on the radial entropy gradient through the
Brunt-V\"ais\"al\"a frequency $N^2 =
-\gamma^{-1}\beta_P\beta_S(H/R)^2\Omega^2$.  This is proportional to
the product of the logarithmic gradients in the pressure $P$ and
entropy $S = \log({P}{\rho^{-\gamma}})$, the latter written as a
potential temperature.  Here the local pressure scale height
$H=c_s/\Omega$ and the gradients are $\beta_P= d{\log}P/d{\log}R$ and
$\beta_S=\beta_T+(1-\gamma) \beta_\rho$, with $\beta_\rho$ the
logarithmic radial density and $\beta_T$ the logarithmic radial
temperature gradient, all for the vertically unstratified case.
Positive $N^2$ corresponds to convective stability and negative to
unstable stratification if the fluid is non-rotating.  In a rotating
system the radial motions are affected if the modified Rayleigh
criterion $(1/R^3)\partial_R[\Omega^2(R)R^4]+N^2>0$ holds \citep[a
  special case of the Solberg-H\o iland criterion with rotation;][]
{Tassoul07, Chandrasekhar61}.  Since typically $-N^2 \ll \Omega^2$,
all radial convective modes are linearly stable \citep{Klahr04,
  Johnson05, Johnson06}.  Yet numerical models indicate vortices
pre-existing in the disk are amplified if thermal relaxation occurs
over approximately the vortices' internal rotation periods
\citep{pjs07, Petersen07b}.
Because growth requires perturbations of finite amplitude,
\citet{lp10} coined the name subcritical baroclinic instability or
SBI.  The vortex growth rate is proportional to $-\tau N^2$, with
$\tau$ the thermal relaxation time, provided $\Omega\tau<1$
\citep{lp10,Raettig13}.  The thermal relaxation times needed for SBI
are generally 100$\times$ longer than the maximum for the GSF in an
isothermal background.  SBI thus seems likely to occur in regions of
greater optical depth.

\begin{figure}[tb!]
\centering
\includegraphics[width=0.94\linewidth]{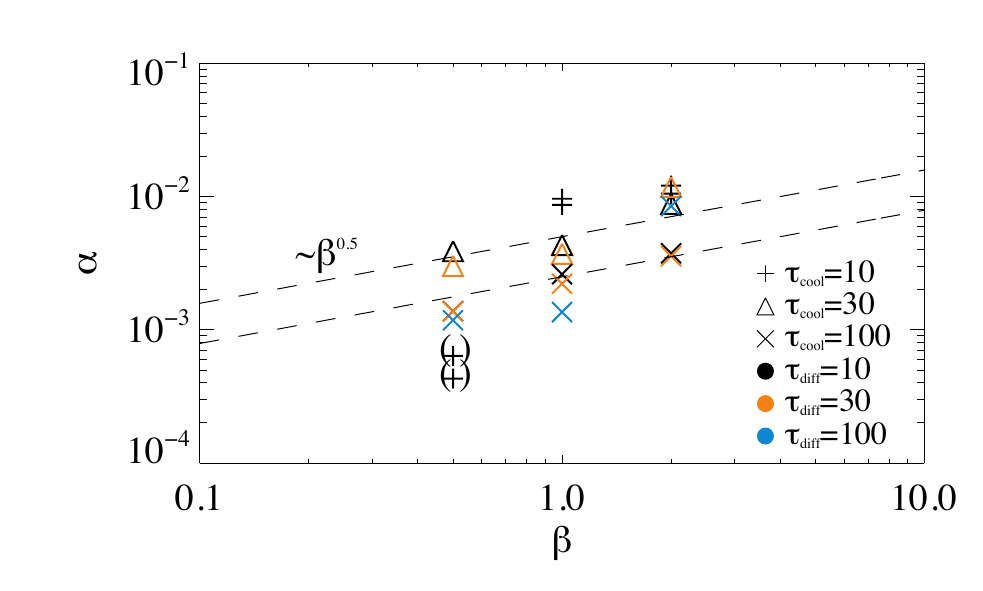}
\caption{Reynolds stress parameter $\alpha$ in saturated 2-D
  baroclinic turbulence vs.\ the logarithmic radial gradient in both
  entropy and pressure, $\beta_S=\beta_P$.  Each calculation is shown
  by a symbol whose color indicates the timescale for thermal
  diffusion within the disk plane, and shape indicates the timescale
  for heating and cooling via the disk surface
  \citep{Raettig13}.  \vspace*{-1.5mm}}
\label{fig-klahr-5}
\end{figure}

Vortices in 3-D are susceptible to the elliptic instability, a
parasitic process \citep{lp09}.  The instability drastically disturbs
the flow in vortices rotating fast enough.  Rotation is faster in
vortices with smaller azimuthal-to-radial aspect ratios, and elliptic
instability typically is important at aspect ratios $<4$.  More
azimuthally-extended vortices may be safe from destruction by the
elliptic instability but still internally unstable, and thus turbulent
at some level.  At least in local 3-D calculations, the parasitic
growth does not completely destroy vortices, but weakens them to
levels where the elliptic instability is suppressed allowing for new
baroclinic growth.  The saturated vortices' turbulent transport
coefficient $\alpha$ is 0.001--0.01 depending on the cooling time and
entropy gradient (Fig.~\ref{fig-klahr-5}), a range compatible with
vortices providing the angular momentum transport behind disks'
observed surface density profiles \citep{2009ApJ...700.1502A,
  2010ApJ...723.1241A}.

\subsection{General Properties of Vortices}

Independently of how they form, vortices extend up to two pressure
scale heights in radius and eight or more in azimuth, in both 2- and
3-D calculations.  Wider vortices experience a supersonic difference
in orbital speed between their inner and outer edges, so a shock
truncates the flow pattern.  Because the structures are large and can
trap solid particles, they might be observable and indeed may already
have been observed \citep{2009ApJ...704..496B, 2012MNRAS.419.1701R,
  2013Sci...340.1199V}.

With magnetic fields well-coupled, vortices are quickly disrupted by a
vortex version of the MRI coined the magneto-elliptic instability
\citep{lk11} and occurring only in 3-D calculations.  Yet if the disk
is magnetically dead this effect should be suppressed.

Like their cousins in the earth's atmosphere, vortices in the body of
the disk are not fixed at a radial or azimuthal location, but tend to
migrate on reaching a certain amplitude as they exchange angular
momentum with their surroundings by emitting spiral waves
\citep{PaardekooperPapaloizou10}.  Consequently vortices are amplified
by the radial buoyancy, then migrate inward and may disappear into the
star or magnetically-active part of the disk.  Thus a reliable
mechanism is needed for initiating vortices and this is yet to be
identified.  Possibilities are a RWI relying on planets or other disk
turbulence, the spiral density waves launched by the previous
generation of vortices, and the pattern generated by the GSF
instability.  Or, perhaps a further hydrodynamic instability triggers
vortex growth.  Vortices could create copies of themselves through a
non-linear ``critical layer'' instability \citep{2013PhRvL.111h4501M}.
This occurs in disks that are stably-stratified in the vertical
direction with little or no thermal relaxation --- diametrically
opposed to the conditions favoring the GSF.  How strict these
requirements are and how well they are realized in protostellar disks
is still unclear.

In summary we emphasize that no hydrodynamical instability is known to
act on a strictly barotropic, vertically unstratified disk without a
local extremum in its rotation profile, nor does any known mechanism
sustain long-lived vortices within \citep{2006ApJ...653..513S}.
However at least two instabilities can operate globally in a
baroclinic disk under suitable thermal conditions: the GSF and
baroclinic vortex driving instabilities.

Motivated by the terrestrial atmospheric weather patterns arising from
individual instabilities driven by solar heating, we have named the
total effect of the hydrodynamical instabilities the ``disk weather''.
A key issue is when and where in protostellar disks the heating and
cooling timescales support such weather.  Calculations to date span
only a small fraction of the parameter space.  For instance there
exists no global, vertically stratified calculation of baroclinic
vortices with realistic entropy gradients and radiation transport,
owing to the high spatial resolution required.  Such calculations will
eventually be needed if we are to study the origin, stability and
ultimate fate of vortices as weather patterns in protostellar disks.

%%%%%%%%%%%%%%%%%%%%%%%%%%%%%%%%%%%%%%%%%%%%%%%%%%%%%%%%%%%%%%%%%%%%%%%%
\section{\textbf{Observational Signatures and Constraints}
  \label{sec:signatures}}
To understand protostellar disks' evolution, we must know which
transport processes are at work.  The many possible mechanisms mean a
reliable answer requires an empirical approach.  Observations probe
the disks' kinematics, density and temperature structure, ionization
state, magnetic fields and composition, each of which carries
signatures of the transport processes.  While few definitive
conclusions can yet be reached, some predictions are now falsifiable.
It is worth remembering that the transport process may differ from one
disk annulus to the next and from one star to another.  For example,
gravitational instability is more likely at the higher surface
densities found in class~I YSOs, while magnetic coupling is easier to
achieve at the lower surface densities found in T~Tauri disks.

\subsection{\textbf{Kinematics}}
Each transport process has a distinct flow pattern that would serve as
a fingerprint if we could only observe with enough spatial and
spectral resolution.  Disk weather makes vortices, gravitational
instability produces spiral shocks, disk winds mean helical outflow
and MRI turbulence is transonic near the disk surface.

Non-thermal velocity dispersions in the outer reaches of a few disks
have been measured using mm and sub-mm lines.  TW~Hya
yielded an upper limit $<0.1 c_s$, while velocities $\sim 0.5 c_s$
were found in the outer disks of HD~163296 \citep{2011ApJ...727...85H}
and DM~Tau \citep{2012A&A...548A..70G}.

\begin{figure}[tb!]
   \centering
   \includegraphics[width=0.9\linewidth]{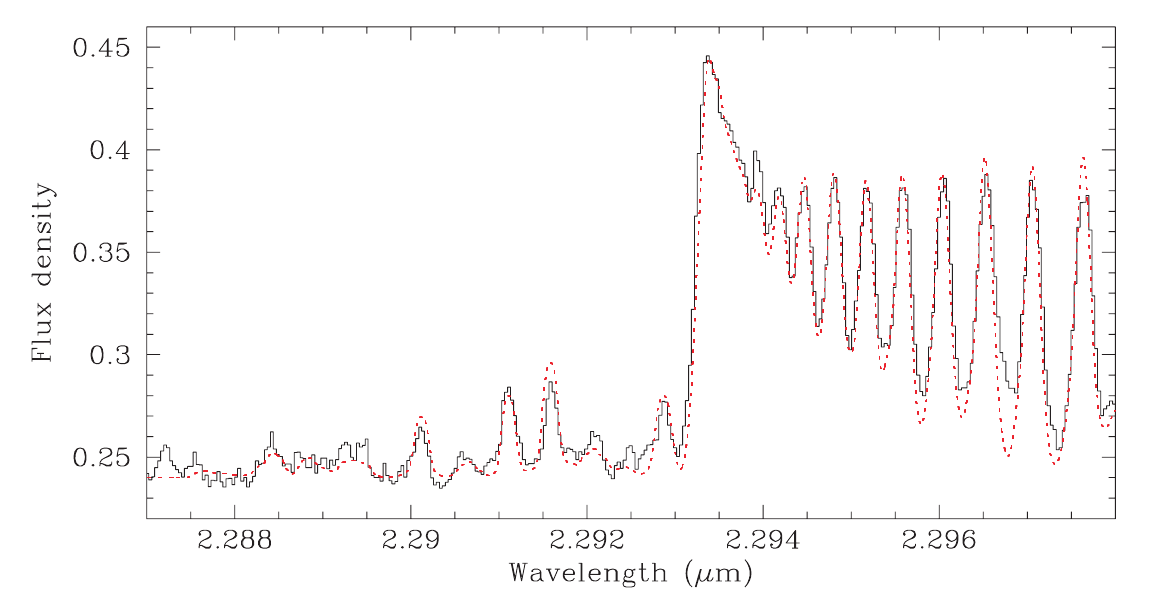}
   \caption{Observed 2.3-$\mu$m spectrum of young star V1331~Cyg
     (histogram) overlaid with a model (dotted).  The microturbulent
     broadening used, 4~km~s$^{-1}$, exceeds the 0.9-km-s$^{-1}$ CO
     thermal dispersion and is near the sound speed at 2500~K
     \citep{2009ApJ...691..738N}.  \vspace*{-1.5mm}  }
   \label{fig:COturbulence}
\end{figure}

Velocities nearer the star are also detected using infrared (IR) lines
formed in the disk atmosphere.  The evidence again is for turbulent
speeds comparable to the sound speed \citep{2004ApJ...603..213C,
  2004ApJ...609..906H, 2009ApJ...691..738N}.  In
Fig.~\ref{fig:COturbulence}, the disk's rotation and microturbulence
are jointly constrained by the widths of the isolated water and CO
lines, the CO lines' beating immediately redward of the bandhead, and
their strengths far from the bandhead compared to the bandhead itself.

Both MRI turbulence \citep{1995ApJ...440..742H, 2000apj...534..398m,
  2011ApJ...743...17S} and gravito-turbulence
\citep{2012MNRAS.426.2419F} have anisotropies that could distinguish
them from other transport processes.  The velocity dispersion
ellipsoid is best accessed using saturated emission lines, especially
those of heavy species for which the thermal speed is less than the
turbulent speed \citep{1995A&A...297..273H}.

Warm disk outflows inferred from atomic oxygen emission line profiles
in the optical \citep{2013ApJ...772...60R} could arise either in a
photoevaporative wind (chapter by Alexander et al.) or in a
magnetocentrifugal wind.  If a cold outflow were observed, it would
uniquely indicate magnetic acceleration.

\subsection{\textbf{Structure in the Gas and Dust}}
Mapping surface densities at mm and sub-mm wavelengths
\citep{2009ApJ...700.1502A, 2009ApJ...701..260I, 2010ApJ...723.1241A}
offers the potential to determine both whether transport mechanisms
can operate and whether they are operating.  Questions of the first
kind include whether the mass column is low enough for ionizing
radiation to reach the midplane and whether the Toomre~Q parameter is
low enough for gravitational instability.  A question of the second
kind, now accessible with ALMA, is whether any spiral features more
closely resemble the density waves raised by self-gravity
\citep{cossins09, cossins10} or those excited in MRI turbulence
\citep{2009MNRAS.397...64H, 2012MNRAS.419.1085H}.

Millimeter mapping also enables detection of vortices.  Modeling
indicates vortices trap solid particles \citep{2006ApJ...649..415I,
  2008A&A...491L..41L, 2009A&A...497..869L, 2012A&A...545A.134M}, and
the continuum emission from mm-sized grains shows big asymmetries in
several cases \citep{2009ApJ...704..496B}.  Among the most spectacular
is a crescent feature observed at wavelength 0.44~mm near the inner
rim of the transitional disk around Oph~IRS~48
\citep{2013Sci...340.1199V}.  The feature is $>$100$\times$ brighter
than emission on the opposite side of the star, while no such
asymmetry appears either in the mid-IR emission from $\mu$m-sized
dust, or in the line emission from CO gas.  Developing such a strong
feature in the particles appears to require a long-lived asymmetry in
the gas \citep{2013A&A...550L...8B, 2013A&A...553L...3A,
  2013ApJ...775...17L}.  Along similar lines, zonal flows will be
detectable by axisymmetric rings and gaps \citep{ruge13}.  Care is
needed to distinguish gaps made by different processes.

\begin{figure}[tb!]
   \centering
   \includegraphics[width=0.9\linewidth]{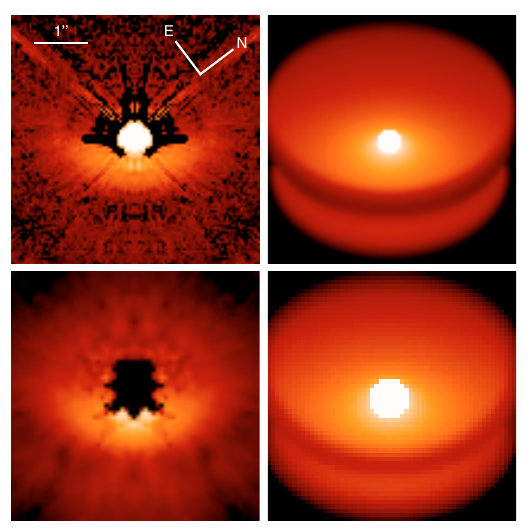}
   \caption{Images of the disk around IM~Lupi at wavelengths 0.6 (top)
     and 1.6~$\mu$m (bottom) from Hubble (left panels), alongside
     models fitted jointly to the images and optical-to-mm-wavelength
     SED (right).  While mm data show substantial grain growth and
     settling, these scattered light images' brightness indicates
     sub-$\mu$m grains remain suspended in the disk atmosphere,
     requiring stirring \citep{2008A&A...489..633P}.  \vspace*{-1.5mm}  }
   \label{fig:scatteredlight}
\end{figure}

Young stars' IR spectral energy distributions (SEDs) and scattered
light imaging sample the distribution of sub-$\mu$m grains.
Turbulence of some kind appears to be needed to keep enough of the
dust suspended in the disk atmospheres to account for the reprocessed
and scattered starlight
\citep[Fig.~\ref{fig:scatteredlight};][]{2007prpl.conf..523W,
  2005a&a...434..971d}.  Typical SEDs of disk-bearing young stars can
be explained using dust stirred by turbulence at 1\% of the sound
speed, almost independent of stellar mass over the range from 0.08 to
2~$M_\odot$ \citep{2012A&A...539A...9M}.  Higher turbulent speeds are
allowed if balanced by reduced dust abundances, or reduced gas masses.

Dust evolution is coupled intimately to turbulence driven by magnetic
forces, because recombination on grains can reduce the gas
conductivity below the threshold for MRI turbulence
\citep{2000apj...543..486s, 2006A&A...445..205I, 2007ap&ss.311...35w}.
In the absence of turbulence, grains grow and settle within a fraction
of the disk lifetime.  On the other hand, turbulent stirring enhances
solid particles' collision rates, speeding growth if the collisions
are slow and causing fragmentation if the collisions are fast
\citep{2008a&a...480..859b, 2010a&a...513a..57z, 2010A&A...513A..79B}.
It remains to be seen whether the mutual coupling can account for the
correlation observed between grain size and settling
\citep{2009ApJS..182..477S} and the lack of correlation of grain size
with other parameters.

Further information on transport processes is contained in the disk
thickness.  MRI turbulence leads to the formation of a
magnetically-supported disk atmosphere that is optically-thin to its
own thermal emission, because good magnetic coupling requires
recombination that is not too fast, and the opacity, like the
recombination, depends on the dust cross-section
\citep{2009apj...701..737b}.  At the same time the magnetized
atmosphere is optically-thick to the starlight entering at a grazing
angle \citep{hirose&turner11}.  The disk thickness then reveals the
magnetic field strengths rather than the internal temperatures.  A
contrary effect comes from magnetic fields generated in a wind with
midplane plasma beta near unity.  These can compress the disk so that
its thickness is less than the hydrostatic value
\citep{2010MNRAS.401..479K}.

Variability in the IR excess is widespread among young stars
with disks, having amplitudes 0.1-0.5~magnitude and timescales as
short as a week \citep{2008ApJ...675.1375L, 2009ApJ...704L..15M,
  2010ApJS..186..111L, 2011ApJ...732...83F}.  In some cases the
variations correlate with markers of the star's accretion and magnetic
activity \citep{2013AJ....145...66F} while in others they do not
\citep{2011ApJ...733...50M}.  The widespread occurrence of the
variability suggests a universal process, such as interaction with the
stellar magnetosphere in the first situation and turbulence-driven
changes in the size or shape of the starlight-reprocessing disk
material in the second.

\subsection{\textbf{Temperature Profiles}}
While accretion flows by their very nature involve the release of
gravitational potential energy, the transport processes described
above differ in where the power is deposited.  A variety of
temperature profiles may result.  Since spectral lines sample the
temperature gradients at the depths where they form, spectroscopy
could be used to constrain which mechanisms are at work, even where it
is not feasible to reach spectral resolutions high enough to measure
kinematics.

The accretion power per unit disk area falls off with distance from
the star as $r^{-3}$ approximately, a steeper dependence than the
illuminating starlight, so that with increasing mass flow rate the
accretion power becomes dominant first near the star.  The accretion
power determines the midplane temperature within 6~AU and the surface
temperature within 2~AU of an $0.9M_\odot$ T~Tauri star accreting at
$10^{-7} M_\odot$~yr$^{-1}$ with the heating distributed in proportion
to the mass \citep{2007prpl.conf..555D}.  Thus the signatures of
accretion heating are most likely to appear in spectral lines formed
within a few AU of the star.

A well-established example is the CO rovibrational first overtone
band, which arises in the disk atmosphere within 1~AU of the star and
can be either in emission, when the atmosphere is hotter than the
interior due to stellar illumination, or in absorption, when the
internal heating is strong enough to dominate
\citep{1991ApJ...380..617C, 2007prpl.conf..507N, 2010AJ....140.1214C}.
The CO rovibrational bands' power source can be the accretion flow if
the associated heating corresponds to a stress approaching the
pressure in the disk atmosphere \citep{2004ApJ...615..972G}.  Stellar
X-ray heating as the power source is insufficient.

MRI turbulence yields dissipation distributed throughout the disk
thickness when the gas is fully-ionized \citep{2010MNRAS.409.1297F,
  2012MNRAS.420.2419F}.  The dissipation occurs higher on average than
the accretion flow, owing to the magnetic fields' buoyancy, and is of
course further concentrated toward the surface if the interior is too
weakly-ionized to couple to the magnetic fields.  Consequently the
midplane is cooler than if the dissipation were distributed uniformly
in the column \citep{hirose&turner11}.  The MRI turbulent layer emits
much of its power in spectral lines.  Line surface brightnesses might
one day reveal the mass flow rate as a function of distance from the
star \citep{2009apj...701..737b}.

\subsection{\textbf{Magnetic Field Measurements}}
To distinguish the two magnetic transport mechanisms --- MRI
turbulence and magnetocentrifugal winds --- it would be valuable to
know how the fields are oriented and whether they are straight or
tangled.  Grains aligned by toroidal magnetic fields in the outer
parts of T~Tauri disks ought to detectably polarize the thermal
emission \citep{2007ApJ...669.1085C} but the sub-mm flux from scales
$>100$~AU is weakly polarized in the five bright protostellar disks
examined so far, with P$<1$\% \citep{2009ApJ...704.1204H,
  2013AJ....145..115H}.  The lack of polarization could be due to
magnetic fields that are either tangled or, for systems viewed
face-on, poloidal rather than toroidal.  Magnetized turbulence and
winds thus cannot yet be separated.  Better sub-mm spatial resolution
and far-IR polarimeters will enable similar tests to be applied to
material nearer the star.

A record of the magnetic fields in the protosolar disk was once
thought to be preserved in the remanent magnetization of chondritic
meteorites last melted during the planet-forming epoch.  The remanence
now appears more likely to sample the magnetic fields in the unmelted
chondritic surface layer of a partially-differentiated planetesimal
that sustained dynamo action in its convecting metallic core
\citep{2010SSRv..152..341W}.

\subsection{\textbf{Detecting Transport Through its Effects on Composition}}

Departures from local chemical equilibrium can indicate material being
transported along gradients, especially gradients in temperature and
radiation intensity.  A classic case is the presence of crystalline
silicate grains in comets, which because of their icy makeup surely
formed beyond the Solar nebula's snow line.  Grains in the
interstellar medium are almost entirely amorphous
\citep{2005ApJ...633..534K}, and crystalline structure is destroyed by
cosmic rays within 70~Myr \citep{2007ApJ...662..372B}.  The comet
grains' crystallinity thus most likely arose by heating to 1000~K or
more within the Solar nebula.  Crystalline silicate grains are
detected in comets through IR spectroscopy \citep{2005Icar..179..158M,
  2006Sci...313..635L} and found in the sample brought back from Comet
81P/Wild~2 by the Stardust mission.  The returned sample shows a wide
range of olivine and pyroxene compositions.  Among the grains is a
refractory, CAI-like particle processed at temperatures above 2000~K
\citep{2006Sci...314.1735Z}.  Taken together, these properties appear
to require large-scale radial transport in the protosolar disk.

Crystalline silicates also occur in the disks around other young
stars.  The silicate grains' crystalline fraction is similar across
the range of annuli in T~Tauri disks producing the 10-33~$\mu$m IR
bands \citep{2009ApJS..182..477S}, indicating either a dispersed
source or efficient transport.  Variations in the 10- and 20-$\mu$m
signatures of crystalline forsterite during an outburst of EX~Lupi
suggest supersonic transport, compatible with grains dragged along by
a wind \citep{2012ApJ...744..118J}.

A new kind of probe of the transport is the mid-IR emission from water
and organic molecules discovered in the atmospheres of many T~Tauri
disks \citep{2008ApJ...676L..49S, 2008Sci...319.1504C,
  2010ApJ...720..887P, 2011ApJ...731..130S, 2011ApJ...733..102C}.  In
particular, there is a trend between the HCN-to-water ratio at AU
distances and the sub-mm continuum flux outside 20~AU, suggested to
arise from water ice accumulating into solid bodies
\citep{2013ApJ...766..134N}.

Disks' compositions can be easier to measure than their kinematics,
but carry fewer clear signatures of specific transport processes since
the details of the flow often matter less than whether the material
has been exposed to high temperatures or energetic radiation.
Nevertheless the transport processes discussed in this chapter have
some distinctive features with the potential to yield differing
chemical signatures:  \vspace*{-1.5mm}

\paragraph{MRI turbulence} yields mixing and angular momentum transfer
coefficients roughly equal \citep{2005ApJ...634.1353J,
  2006apj...639.1218t, 2006a&a...452..751f} with the angular momentum
transfer increasing faster than the mixing as the background vertical
magnetic field is made stronger \citep{2006MNRAS.370L..71J}.  As
discussed in \S\ref{sec:linearmri} the magnetic activity at
planet-forming distances is strongest in the disk surface layers.  The
global meridional circulation predicted in simple viscous models has
not so far been seen in MHD calculations \citep{fromangetal11,
  flocketal11}.  \vspace*{-1.5mm}

\paragraph{Magnetocentrifugal winds} drive the gas radially inward
near the midplane and outward above the height where the inclined
field lines' azimuthal speed reaches Keplerian, in radially-localized
stationary solutions \citep{skw11}.  Rapid inward and outward radial
transport near the surface of a wind-driving disk could also come from
episodic channel flows \citep{si09} whose repeated breakup would cause
some vertical mixing.  \vspace*{-1.5mm}

\paragraph{Gravitoturbulence} can redistribute material in radius as
rapidly as can MRI turbulence \citep{Michael12}.  Even so a
contaminant may remain inhomogeneous during the global mixing
timescale if the unstable features are bigger than the disk thickness
\citep{2007ApJ...660.1707B, 2013ApJ...773....5B} or if shock passages
affect desorption from grains or destruction in chemical reactions
\citep{2011MNRAS.417.2950I}.  The shock heating associated with
gravitational instability appears to drive little convection in the
vertical direction when an accurate boundary condition is applied at
the disk photosphere \citep{2010ApJ...716L.176C}.  \vspace*{-1.5mm}

\paragraph{}
Several attempts have been made to bring these ideas about transport
into contact with composition measurements.  Crystalline grains can be
moved through the disk to the comet-forming region with either a
sustained radially-outward midplane flow, or a diffusivity for
turbulent mixing exceeding that for angular momentum transfer
\citep{2010ApJ...719.1633H}.  Vertical mixing with a coefficient in
the range inferred for the radial angular momentum transfer enhances
the abundances of some carbon- and sulfur-bearing molecules in T~Tauri
disks by orders of magnitude, by bringing reactive atoms and ions from
the overlying photodissociation layers into more shielded molecular
regions \citep{2006ApJ...644.1202W, 2011apjs..196...25s}.  The
molecular abundances are affected more by turbulent mixing than by a
steady-state disk wind \citep{2011ApJ...731..115H}.  Comparing these
models with the molecular column densities obtained from mm line
measurements, mostly near and outside 100~AU, yields mixed results
suggesting further complexities in the chemistry, dynamics or both.

%%%%%%%%%%%%%%%%%%%%%%%%%%%%%%%%%%%%%%%%%%%%%%%%%%%%%%%%%%%%%%%%%%%%%%%%
\section{\textbf{Summary and Outlook}
  \label{sec:outlook}}
Magneto-rotational turbulence is suppressed by ambipolar diffusion in
local stratified models of T~Tauri disks' planet-forming regions, even
when the Elsasser number and plasma beta criteria indicate turbulence
in a surface layer about one scale height thick \citep{bs13}.
Furthermore the turbulence is suppressed by a strong Hall effect in
unstratified calculations \citep{2013MNRAS.434.2295K}.  These results
raise serious doubts about whether MRI turbulence is effective at
1--10~AU even in surface layers.  Further progress might involve (1)
bringing the temperatures in MHD calculations up to the level of
sophistication found in thermo-chemical models, where stellar X-ray
and UV photons heat the disk atmosphere to $>$2000~K
\citep{2004ApJ...615..972G, 2008ApJ...688..398E, 2010MNRAS.401.1415O,
  2011A&A...526A.163A, 2013ApJ...766....8A}, surely affecting the
conductivity, and (2) allowing the Ohmic, Hall and ambipolar terms all
to vary with height.  No non-linear MHD calculation yet treats this
full problem.  MRI turbulence likely remains important in the
thermally-ionized gas inside 1~AU and the less-dense material beyond
10~AU.  A key unresolved issue here is the saturation amplitudes at
realistic Prandtl numbers.

Magnetocentrifugal disk winds appear to be favored in disks with the
interior dominated by Ohmic diffusion and the surface dominated by
ambipolar diffusion, threaded by magnetic fields that are not too
weak.  Disk winds and MRI turbulence coexist if the magnetic fields
are weak enough that the MRI wavelength fits in the disk thickness.
The dependences of both MRI turbulence and magnetocentrifugal winds on
the net field point to the large-scale magnetic flux transport as a
major issue that must be adequately dealt if we are to understand
protostellar disks' long-term evolution.

Gravitational instability occurs in disks whose mass ratios with their
central stars exceed their aspect ratios.  It is most easily triggered
in the outer reaches, since $c_s\Omega$ falls off faster with radius
than the typical surface density profile.  Gravitational instability
can be important during the protostellar phase when the disk is fed
rapidly with material falling in from the surrounding envelope.  The
outcome of the instability depends on how quickly the compressed gas
cools, and room remains for dynamical modeling with more complete
treatment of the radiative losses.

All the disk weather processes discussed here, including Rossby and
baroclinic vortices and the Goldreich-Schubert-Fricke instability,
depend on radiative forcing for their energy supply.  There is an
urgent need for calculations with better heating and cooling.  The
instabilities grow slower than the MRI and so can reach significant
amplitudes only in regions lacking MRI turbulence.  Whether they
survive in the presence of a magnetically-launched wind is unclear.

While all the transport processes we discuss are commonly measured and
compared using the Shakura-Sunyaev $\alpha$-parameter, none behaves
like the simple \citet{1973a&a....24..337s} model.  MRI turbulent
stresses fall off with height slower than the gas pressure, even when
the magnetic coupling is good throughout.  Magnetocentrifugal winds
expel the angular momentum vertically rather than radially.  The
energy escapes in a Poynting flux that dissipates far away, leaving
the disk unheated.  Gravitational instability leads to intermittent
shock dissipation, and disk weather yields transport localized around
vortices.  Any treatment of these processes using an $\alpha$-model
should therefore be evaluated carefully.

Observations offer tantalizing clues to the transport processes.
Non-thermal motions are detected, but the spatial resolution is too
coarse to unequivocally distinguish turbulence from winds or vortices.
In order of decreasing distance from the central star,
\begin{itemize}
\item Millimeter lines formed outside $\sim 100$~AU in two disks show
  transonic superthermal widths \citep{2011ApJ...727...85H,
    2012A&A...548A..70G}.  These are explained most easily using
  turbulence but could also arise from outflows.
\item Keeping enough dust suspended in the atmosphere to explain both
  the scattered starlight at $\sim 50$~AU and the SEDs requires some
  kind of stirring.  MRI turbulence can do the job where magnetic
  coupling is good.  Outflows and vortices could potentially also
  reproduce the dust distributions, but better models are called for.
\item Asymmetric sub-mm surface brightness maps in a few disks
  \citep{2013Sci...340.1199V} have so far been reproduced only using
  long-lived non-axisymmetric gas density disturbances.  Vortices are
  the most viable explanation.
\item Applying angular momentum constraints to jets yields launch
  points within a few AU of the stars \citep{2003ApJ...590L.107A,
    2007ApJ...663..350C}.  Disks' inner regions clearly launch
  outflows.
\item Superthermal CO overtone linewidths indicate transonic
  turbulence in the atmosphere inside 1~AU, where the gas is thermally
  ionized, consistent with MRI \citep{2004ApJ...603..213C,
    2004ApJ...609..906H, 2009ApJ...691..738N}.  On the other hand, the
  blueshifted central peaks in many young stars' CO fundamental band
  suggest outflows \citep{2013ApJ...770...94B}.  The two kinds of
  flows likely occur together in some disks.
\end{itemize}
A key to further progress in understanding protostellar disk evolution
is to bring the models closer to observations like these, seeking
clearer signatures of the proposed transport processes.
\vspace*{-1.5mm}

\acknowledgements NT was supported at the Jet Propulsion Laboratory,
California Institute of Technology by NASA grant 11-OSS11-0074; SF by
the European Research Council under the E.U. Seventh Framework
Programme (FP7/2007-2013) / ERC Grant agreement 258729; HK by the
Deutsche For\-schungs\-ge\-mein\-schaft Schwer\-punkt\-pro\-gramm (DFG
SPP) 1385 ``The first ten million years of the solar system''; GL by
the European Community via contract PCIG09-GA-2011-294110; and MW by
Australian Research Council grant DP120101792.
\vspace*{-1.5mm}

%%%%%%%%%%%%%%%%%%%%%%%%%%%%%%%%%%%%%%%%%%%%%%%%%%%%%%%%%%%%%%%%%%%%%%%%
\bigskip

\bibliographystyle{ppvi_lim1}
\bibliography{disktranspcut}

\begin{thebibliography}{312}
\parskip=0pt \itemsep=0pt \small \baselineskip=11pt
\providecommand{\natexlab}[1]{#1}

\bibitem[\protect\astroncite{\emph{{Akimkin}
  et~al.}}{2013}]{2013ApJ...766....8A}
{Akimkin} V. et~al. (2013) \emph{\apj}, \emph{766}, 8.

\bibitem[\protect\astroncite{\emph{{Anderson}
  et~al.}}{2003}]{2003ApJ...590L.107A}
{Anderson} J.~M. et~al. (2003) \emph{\apjl}, \emph{590}, L107.

\bibitem[\protect\astroncite{\emph{{Andrews}
  et~al.}}{2009}]{2009ApJ...700.1502A}
{Andrews} S.~M. et~al. (2009) \emph{\apj}, \emph{700}, 1502.

\bibitem[\protect\astroncite{\emph{{Andrews}
  et~al.}}{2010}]{2010ApJ...723.1241A}
{Andrews} S.~M. et~al. (2010) \emph{\apj}, \emph{723}, 1241.

\bibitem[\protect\astroncite{\emph{{Aresu} et~al.}}{2011}]{2011A&A...526A.163A}
{Aresu} G. et~al. (2011) \emph{\aap}, \emph{526}, A163.

\bibitem[\protect\astroncite{\emph{{Arlt} and
  {Urpin}}}{2004}]{2004A&A...426..755A}
{Arlt} R. and {Urpin} V. (2004) \emph{\aap}, \emph{426}, 755.

\bibitem[\protect\astroncite{\emph{{Ataiee}
  et~al.}}{2013}]{2013A&A...553L...3A}
{Ataiee} S. et~al. (2013) \emph{\aap}, \emph{553}, L3.

\bibitem[\protect\astroncite{\emph{{Avila}}}{2012}]{Avila12}
{Avila} M. (2012) \emph{Physical Review Letters}, \emph{108}, 12, 124501.

\bibitem[\protect\astroncite{\emph{{Bai}}}{2011{\natexlab{a}}}]{2011ApJ...739...50B}
{Bai} X.-N. (2011{\natexlab{a}}) \emph{\apj}, \emph{739}, 50.

\bibitem[\protect\astroncite{\emph{{Bai}}}{2011{\natexlab{b}}}]{2011ApJ...739...51B}
{Bai} X.-N. (2011{\natexlab{b}}) \emph{\apj}, \emph{739}, 51.

\bibitem[\protect\astroncite{\emph{{Bai}}}{2013}]{2013ApJ...772...96B}
{Bai} X.-N. (2013) \emph{\apj}, \emph{772}, 96.

\bibitem[\protect\astroncite{\emph{{Bai} and
  {Goodman}}}{2009}]{2009apj...701..737b}
{Bai} X.-N. and {Goodman} J. (2009) \emph{\apj}, \emph{701}, 737.

\bibitem[\protect\astroncite{\emph{{Bai} and
  {Stone}}}{2011}]{2011apj...736..144b}
{Bai} X.-N. and {Stone} J.~M. (2011) \emph{\apj}, \emph{736}, 144.

\bibitem[\protect\astroncite{\emph{{Bai} and
  {Stone}}}{2013{\natexlab{a}}}]{bs12}
{Bai} X.-N. and {Stone} J.~M. (2013{\natexlab{a}}) \emph{\apj}, \emph{767}, 30.

\bibitem[\protect\astroncite{\emph{{Bai} and
  {Stone}}}{2013{\natexlab{b}}}]{bs13}
{Bai} X.-N. and {Stone} J.~M. (2013{\natexlab{b}}) \emph{\apj}, \emph{769}, 76.

\bibitem[\protect\astroncite{\emph{{Balbus} and
  {Hawley}}}{1991}]{1991ApJ...376..214B}
{Balbus} S.~A. and {Hawley} J.~F. (1991) \emph{\apj}, \emph{376}, 214.

\bibitem[\protect\astroncite{\emph{{Balbus} and
  {Henri}}}{2008}]{2008ApJ...674..408B}
{Balbus} S.~A. and {Henri} P. (2008) \emph{\apj}, \emph{674}, 408.

\bibitem[\protect\astroncite{\emph{{Balbus} and
  {Papaloizou}}}{1999}]{BalbPap99}
{Balbus} S.~A. and {Papaloizou} J.~C.~B. (1999) \emph{\apj}, \emph{521}, 650.

\bibitem[\protect\astroncite{\emph{{Balmforth} and
  {Korycansky}}}{2001}]{Balmforth01}
{Balmforth} N.~J. and {Korycansky} D.~G. (2001) \emph{\mnras}, \emph{326}, 833.

\bibitem[\protect\astroncite{\emph{{Barge} and
  {Sommeria}}}{1995}]{BargeSommeria95}
{Barge} P. and {Sommeria} J. (1995) \emph{\aap}, \emph{295}, L1.

\bibitem[\protect\astroncite{\emph{{Barranco}}}{2009}]{2009ApJ...691..907B}
{Barranco} J.~A. (2009) \emph{\apj}, \emph{691}, 907.

\bibitem[\protect\astroncite{\emph{{Beckwith} et~al.}}{2009}]{bhk09}
{Beckwith} K. et~al. (2009) \emph{\apj}, \emph{707}, 428.

\bibitem[\protect\astroncite{\emph{{Beckwith}
  et~al.}}{2011}]{2011MNRAS.416..361B}
{Beckwith} K. et~al. (2011) \emph{\mnras}, \emph{416}, 361.

\bibitem[\protect\astroncite{\emph{{Bergin}
  et~al.}}{2007}]{2007prpl.conf..751B}
{Bergin} E.~A. et~al. (2007) \emph{Protostars and Planets V}, pp. 751--766.

\bibitem[\protect\astroncite{\emph{{Bethell} and
  {Bergin}}}{2011}]{2011ApJ...739...78B}
{Bethell} T.~J. and {Bergin} E.~A. (2011) \emph{\apj}, \emph{739}, 78.

\bibitem[\protect\astroncite{\emph{{Birnstiel}
  et~al.}}{2010}]{2010A&A...513A..79B}
{Birnstiel} T. et~al. (2010) \emph{\aap}, \emph{513}, A79.

\bibitem[\protect\astroncite{\emph{{Birnstiel}
  et~al.}}{2013}]{2013A&A...550L...8B}
{Birnstiel} T. et~al. (2013) \emph{\aap}, \emph{550}, L8.

\bibitem[\protect\astroncite{\emph{{Blaes} and
  {Balbus}}}{1994}]{1994ApJ...421..163B}
{Blaes} O.~M. and {Balbus} S.~A. (1994) \emph{\apj}, \emph{421}, 163.

\bibitem[\protect\astroncite{\emph{{Blandford} and {Payne}}}{1982}]{BP82}
{Blandford} R.~D. and {Payne} D.~G. (1982) \emph{\mnras}, \emph{199}, 883.

\bibitem[\protect\astroncite{\emph{{Bodo} et~al.}}{2008}]{2008A&A...487....1B}
{Bodo} G. et~al. (2008) \emph{\aap}, \emph{487}, 1.

\bibitem[\protect\astroncite{\emph{{Bodo} et~al.}}{2011}]{bodoetal11}
{Bodo} G. et~al. (2011) \emph{\apj}, \emph{739}, 82.

\bibitem[\protect\astroncite{\emph{{Boley}}}{2009}]{Boley09}
{Boley} A.~C. (2009) \emph{\apjl}, \emph{695}, L53.

\bibitem[\protect\astroncite{\emph{{Boss}}}{1997}]{Boss1997}
{Boss} A.~P. (1997) \emph{Science}, \emph{276}, 1836.

\bibitem[\protect\astroncite{\emph{{Boss}}}{2007{\natexlab{a}}}]{2007ApJ...660.1707B}
{Boss} A.~P. (2007{\natexlab{a}}) \emph{\apj}, \emph{660}, 1707.

\bibitem[\protect\astroncite{\emph{{Boss}}}{2007{\natexlab{b}}}]{Boss07}
{Boss} A.~P. (2007{\natexlab{b}}) \emph{\apjl}, \emph{661}, L73.

\bibitem[\protect\astroncite{\emph{{Boss}}}{2013}]{2013ApJ...773....5B}
{Boss} A.~P. (2013) \emph{\apj}, \emph{773}, 5.

\bibitem[\protect\astroncite{\emph{{Brandenburg}
  et~al.}}{1995}]{1995ApJ...446..741B}
{Brandenburg} A. et~al. (1995) \emph{\apj}, \emph{446}, 741.

\bibitem[\protect\astroncite{\emph{{Brauer}
  et~al.}}{2008}]{2008a&a...480..859b}
{Brauer} F. et~al. (2008) \emph{\aap}, \emph{480}, 859.

\bibitem[\protect\astroncite{\emph{{Bringa}
  et~al.}}{2007}]{2007ApJ...662..372B}
{Bringa} E.~M. et~al. (2007) \emph{\apj}, \emph{662}, 372.

\bibitem[\protect\astroncite{\emph{{Brown} et~al.}}{2009}]{2009ApJ...704..496B}
{Brown} J.~M. et~al. (2009) \emph{\apj}, \emph{704}, 496.

\bibitem[\protect\astroncite{\emph{{Brown} et~al.}}{2013}]{2013ApJ...770...94B}
{Brown} J.~M. et~al. (2013) \emph{\apj}, \emph{770}, 94.

\bibitem[\protect\astroncite{\emph{{Cai} et~al.}}{2008}]{Cai08}
{Cai} K. et~al. (2008) \emph{\apj}, \emph{673}, 1138.

\bibitem[\protect\astroncite{\emph{{Cai} et~al.}}{2010}]{2010ApJ...716L.176C}
{Cai} K. et~al. (2010) \emph{\apjl}, \emph{716}, L176.

\bibitem[\protect\astroncite{\emph{{Calvet}
  et~al.}}{1991}]{1991ApJ...380..617C}
{Calvet} N. et~al. (1991) \emph{\apj}, \emph{380}, 617.

\bibitem[\protect\astroncite{\emph{{Cao} and {Spruit}}}{2002}]{cs02}
{Cao} X. and {Spruit} H.~C. (2002) \emph{\aap}, \emph{385}, 289.

\bibitem[\protect\astroncite{\emph{{Carr} and
  {Najita}}}{2008}]{2008Sci...319.1504C}
{Carr} J.~S. and {Najita} J.~R. (2008) \emph{Science}, \emph{319}, 1504.

\bibitem[\protect\astroncite{\emph{{Carr} and
  {Najita}}}{2011}]{2011ApJ...733..102C}
{Carr} J.~S. and {Najita} J.~R. (2011) \emph{\apj}, \emph{733}, 102.

\bibitem[\protect\astroncite{\emph{{Carr} et~al.}}{2004}]{2004ApJ...603..213C}
{Carr} J.~S. et~al. (2004) \emph{\apj}, \emph{603}, 213.

\bibitem[\protect\astroncite{\emph{{Chandrasekhar}}}{1960}]{1960PNAS...46..253C}
{Chandrasekhar} S. (1960) \emph{Proceedings of the National Academy of
  Science}, \emph{46}, 253.

\bibitem[\protect\astroncite{\emph{{Chandrasekhar}}}{1961}]{Chandrasekhar61}
{Chandrasekhar} S. (1961) \emph{{Hydrodynamic and hydromagnetic stability}},
  Dover Publications; Dover edition (February 1, 1981).

\bibitem[\protect\astroncite{\emph{{Cho} and
  {Lazarian}}}{2007}]{2007ApJ...669.1085C}
{Cho} J. and {Lazarian} A. (2007) \emph{\apj}, \emph{669}, 1085.

\bibitem[\protect\astroncite{\emph{{Ciesla} and
  {Cuzzi}}}{2006}]{2006Icar..181..178C}
{Ciesla} F.~J. and {Cuzzi} J.~N. (2006) \emph{\icarus}, \emph{181}, 178.

\bibitem[\protect\astroncite{\emph{{Clarke} et~al.}}{2007}]{Clarke07}
{Clarke} C.~J. et~al. (2007) \emph{\mnras}, \emph{381}, 1543.

\bibitem[\protect\astroncite{\emph{{Cleeves}
  et~al.}}{2013}]{2013ApJ...772....5I}
{Cleeves} L.~I. et~al. (2013) \emph{\apj}, \emph{772}, 5.

\bibitem[\protect\astroncite{\emph{{Coffey}
  et~al.}}{2007}]{2007ApJ...663..350C}
{Coffey} D. et~al. (2007) \emph{\apj}, \emph{663}, 350.

\bibitem[\protect\astroncite{\emph{{Connelley} and
  {Greene}}}{2010}]{2010AJ....140.1214C}
{Connelley} M.~S. and {Greene} T.~P. (2010) \emph{\aj}, \emph{140}, 1214.

\bibitem[\protect\astroncite{\emph{{Cossins} et~al.}}{2009}]{cossins09}
{Cossins} P. et~al. (2009) \emph{\mnras}, \emph{393}, 1157.

\bibitem[\protect\astroncite{\emph{{Cossins} et~al.}}{2010}]{cossins10}
{Cossins} P. et~al. (2010) \emph{\mnras}, \emph{407}, 181.

\bibitem[\protect\astroncite{\emph{{Davis} et~al.}}{2010}]{davisetal10}
{Davis} S.~W. et~al. (2010) \emph{\apj}, \emph{713}, 52.

\bibitem[\protect\astroncite{\emph{{de Val-Borro}
  et~al.}}{2007}]{deVal-Borro07}
{de Val-Borro} M. et~al. (2007) \emph{\aap}, \emph{471}, 1043.

\bibitem[\protect\astroncite{\emph{{Desch}}}{2004}]{2004ApJ...608..509D}
{Desch} S.~J. (2004) \emph{\apj}, \emph{608}, 509.

\bibitem[\protect\astroncite{\emph{{Desch} et~al.}}{2004}]{2004ApJ...602..528D}
{Desch} S.~J. et~al. (2004) \emph{\apj}, \emph{602}, 528.

\bibitem[\protect\astroncite{\emph{{Dittrich}
  et~al.}}{2013}]{2013ApJ...763..117D}
{Dittrich} K. et~al. (2013) \emph{\apj}, \emph{763}, 117.

\bibitem[\protect\astroncite{\emph{{Douglas} et~al.}}{2013}]{douglas2013}
{Douglas} T.~A. et~al. (2013) \emph{\mnras}, \emph{433}, 2064.

\bibitem[\protect\astroncite{\emph{{Dullemond} and
  {Dominik}}}{2005}]{2005a&a...434..971d}
{Dullemond} C.~P. and {Dominik} C. (2005) \emph{\aap}, \emph{434}, 971.

\bibitem[\protect\astroncite{\emph{{Dullemond}
  et~al.}}{2007}]{2007prpl.conf..555D}
{Dullemond} C.~P. et~al. (2007) \emph{Protostars and Planets V}, pp. 555--572.

\bibitem[\protect\astroncite{\emph{{Dzyurkevich}
  et~al.}}{2010}]{2010A&A...515A..70D}
{Dzyurkevich} N. et~al. (2010) \emph{\aap}, \emph{515}, A70.

\bibitem[\protect\astroncite{\emph{{Dzyurkevich}
  et~al.}}{2013}]{2013ApJ...765..114D}
{Dzyurkevich} N. et~al. (2013) \emph{\apj}, \emph{765}, 114.

\bibitem[\protect\astroncite{\emph{{Ercolano}
  et~al.}}{2008}]{2008ApJ...688..398E}
{Ercolano} B. et~al. (2008) \emph{\apj}, \emph{688}, 398.

\bibitem[\protect\astroncite{\emph{{Ferreira}}}{1997}]{f97}
{Ferreira} J. (1997) \emph{\aap}, \emph{319}, 340.

\bibitem[\protect\astroncite{\emph{{Ferreira} and {Pelletier}}}{1993}]{fp93}
{Ferreira} J. and {Pelletier} G. (1993) \emph{\aap}, \emph{276}, 625.

\bibitem[\protect\astroncite{\emph{{Ferreira} and {Pelletier}}}{1995}]{fp95}
{Ferreira} J. and {Pelletier} G. (1995) \emph{\aap}, \emph{295}, 807.

\bibitem[\protect\astroncite{\emph{{Ferro-Font{\'a}n} and {G{\'o}mez de
  Castro}}}{2003}]{2003MNRAS.342..427F}
{Ferro-Font{\'a}n} C. and {G{\'o}mez de Castro} A.~I. (2003) \emph{\mnras},
  \emph{342}, 427.

\bibitem[\protect\astroncite{\emph{{Flaherty}
  et~al.}}{2011}]{2011ApJ...732...83F}
{Flaherty} K.~M. et~al. (2011) \emph{\apj}, \emph{732}, 83.

\bibitem[\protect\astroncite{\emph{{Flaherty}
  et~al.}}{2013}]{2013AJ....145...66F}
{Flaherty} K.~M. et~al. (2013) \emph{\aj}, \emph{145}, 66.

\bibitem[\protect\astroncite{\emph{{Flaig} et~al.}}{2010}]{2010MNRAS.409.1297F}
{Flaig} M. et~al. (2010) \emph{\mnras}, \emph{409}, 1297.

\bibitem[\protect\astroncite{\emph{{Flaig} et~al.}}{2012}]{2012MNRAS.420.2419F}
{Flaig} M. et~al. (2012) \emph{\mnras}, \emph{420}, 2419.

\bibitem[\protect\astroncite{\emph{{Fleming} and
  {Stone}}}{2003}]{fleming&stone03}
{Fleming} T. and {Stone} J.~M. (2003) \emph{\apj}, \emph{585}, 908.

\bibitem[\protect\astroncite{\emph{{Flock} et~al.}}{2011}]{flocketal11}
{Flock} M. et~al. (2011) \emph{\apj}, \emph{735}, 122.

\bibitem[\protect\astroncite{\emph{{Flock} et~al.}}{2012}]{flocketal12}
{Flock} M. et~al. (2012) \emph{\apj}, \emph{744}, 144.

\bibitem[\protect\astroncite{\emph{{Forgan} and {Rice}}}{2013}]{Forgan13}
{Forgan} D. and {Rice} K. (2013) \emph{\mnras}, \emph{430}, 2082.

\bibitem[\protect\astroncite{\emph{{Forgan} et~al.}}{2011}]{Forgan11}
{Forgan} D. et~al. (2011) \emph{\mnras}, \emph{410}, 994.

\bibitem[\protect\astroncite{\emph{{Forgan}
  et~al.}}{2012}]{2012MNRAS.426.2419F}
{Forgan} D. et~al. (2012) \emph{\mnras}, \emph{426}, 2419.

\bibitem[\protect\astroncite{\emph{{Fricke}}}{1968}]{Fricke68}
{Fricke} K. (1968) \emph{\zap}, \emph{68}, 317.

\bibitem[\protect\astroncite{\emph{{Fromang}}}{2005}]{Fromang2005}
{Fromang} S. (2005) \emph{\aap}, \emph{441}, 1.

\bibitem[\protect\astroncite{\emph{{Fromang} and
  {Nelson}}}{2006}]{fromang&nelson06}
{Fromang} S. and {Nelson} R.~P. (2006) \emph{\aap}, \emph{457}, 343.

\bibitem[\protect\astroncite{\emph{{Fromang} and
  {Nelson}}}{2009}]{fromang&nelson09}
{Fromang} S. and {Nelson} R.~P. (2009) \emph{\aap}, \emph{496}, 597.

\bibitem[\protect\astroncite{\emph{{Fromang} and
  {Papaloizou}}}{2006}]{2006a&a...452..751f}
{Fromang} S. and {Papaloizou} J. (2006) \emph{\aap}, \emph{452}, 751.

\bibitem[\protect\astroncite{\emph{{Fromang} and
  {Papaloizou}}}{2007}]{fromang&pap07}
{Fromang} S. and {Papaloizou} J. (2007) \emph{\aap}, \emph{476}, 1113.

\bibitem[\protect\astroncite{\emph{{Fromang} et~al.}}{2007}]{fromangetal07}
{Fromang} S. et~al. (2007) \emph{\aap}, \emph{476}, 1123.

\bibitem[\protect\astroncite{\emph{{Fromang} et~al.}}{2011}]{fromangetal11}
{Fromang} S. et~al. (2011) \emph{\aap}, \emph{534}, A107.

\bibitem[\protect\astroncite{\emph{{Fromang} et~al.}}{2013}]{fllo13}
{Fromang} S. et~al. (2013) \emph{\aap}, \emph{552}, A71.

\bibitem[\protect\astroncite{\emph{{Gammie}}}{1996}]{1996apj...457..355g}
{Gammie} C.~F. (1996) \emph{\apj}, \emph{457}, 355.

\bibitem[\protect\astroncite{\emph{{Gammie}}}{2001}]{Gam01}
{Gammie} C.~F. (2001) \emph{\apj}, \emph{553}, 174.

\bibitem[\protect\astroncite{\emph{{Garmire}
  et~al.}}{2000}]{2000aj....120.1426g}
{Garmire} G. et~al. (2000) \emph{\aj}, \emph{120}, 1426.

\bibitem[\protect\astroncite{\emph{{Glassgold}
  et~al.}}{2004}]{2004ApJ...615..972G}
{Glassgold} A.~E. et~al. (2004) \emph{\apj}, \emph{615}, 972.

\bibitem[\protect\astroncite{\emph{{Gnedin} et~al.}}{1995}]{gnedin1995}
{Gnedin} O.~Y. et~al. (1995) \emph{\aj}, \emph{110}, 1105.

\bibitem[\protect\astroncite{\emph{{Goldreich} and
  {Lynden-Bell}}}{1965}]{GoldBell65}
{Goldreich} P. and {Lynden-Bell} D. (1965) \emph{\mnras}, \emph{130}, 125.

\bibitem[\protect\astroncite{\emph{{Goldreich} and
  {Schubert}}}{1967}]{Goldreich67}
{Goldreich} P. and {Schubert} G. (1967) \emph{\apj}, \emph{150}, 571.

\bibitem[\protect\astroncite{\emph{{Gressel}}}{2010}]{gressel10}
{Gressel} O. (2010) \emph{\mnras}, \emph{405}, 41.

\bibitem[\protect\astroncite{\emph{{Gressel}
  et~al.}}{2011}]{2011MNRAS.415.3291G}
{Gressel} O. et~al. (2011) \emph{\mnras}, \emph{415}, 3291.

\bibitem[\protect\astroncite{\emph{{Guan} et~al.}}{2009}]{guanetal09}
{Guan} X. et~al. (2009) \emph{\apj}, \emph{694}, 1010.

\bibitem[\protect\astroncite{\emph{{Guilloteau}
  et~al.}}{2012}]{2012A&A...548A..70G}
{Guilloteau} S. et~al. (2012) \emph{\aap}, \emph{548}, A70.

\bibitem[\protect\astroncite{\emph{{Harsono} et~al.}}{2011}]{Harsono12}
{Harsono} D. et~al. (2011) \emph{\mnras}, \emph{413}, 423.

\bibitem[\protect\astroncite{\emph{{Hartmann}
  et~al.}}{2004}]{2004ApJ...609..906H}
{Hartmann} L. et~al. (2004) \emph{\apj}, \emph{609}, 906.

\bibitem[\protect\astroncite{\emph{{Hawley} and
  {Balbus}}}{1992}]{1992ApJ...400..595H}
{Hawley} J.~F. and {Balbus} S.~A. (1992) \emph{\apj}, \emph{400}, 595.

\bibitem[\protect\astroncite{\emph{{Hawley} and
  {Stone}}}{1995}]{1995CoPhC..89..127H}
{Hawley} J.~F. and {Stone} J.~M. (1995) \emph{Computer Physics Communications},
  \emph{89}, 127.

\bibitem[\protect\astroncite{\emph{{Hawley} and
  {Stone}}}{1998}]{1998ApJ...501..758H}
{Hawley} J.~F. and {Stone} J.~M. (1998) \emph{\apj}, \emph{501}, 758.

\bibitem[\protect\astroncite{\emph{{Hawley}
  et~al.}}{1995}]{1995ApJ...440..742H}
{Hawley} J.~F. et~al. (1995) \emph{\apj}, \emph{440}, 742.

\bibitem[\protect\astroncite{\emph{{Hawley} et~al.}}{1996}]{hgb96}
{Hawley} J.~F. et~al. (1996) \emph{\apj}, \emph{464}, 690.

\bibitem[\protect\astroncite{\emph{{Hawley}
  et~al.}}{2013}]{2013ApJ...772..102H}
{Hawley} J.~F. et~al. (2013) \emph{\apj}, \emph{772}, 102.

\bibitem[\protect\astroncite{\emph{{Heinemann} and
  {Papaloizou}}}{2009}]{2009MNRAS.397...64H}
{Heinemann} T. and {Papaloizou} J.~C.~B. (2009) \emph{\mnras}, \emph{397}, 64.

\bibitem[\protect\astroncite{\emph{{Heinemann} and
  {Papaloizou}}}{2012}]{2012MNRAS.419.1085H}
{Heinemann} T. and {Papaloizou} J.~C.~B. (2012) \emph{\mnras}, \emph{419},
  1085.

\bibitem[\protect\astroncite{\emph{{Heinzeller}
  et~al.}}{2011}]{2011ApJ...731..115H}
{Heinzeller} D. et~al. (2011) \emph{\apj}, \emph{731}, 115.

\bibitem[\protect\astroncite{\emph{Herault et~al.}}{2011}]{HR11}
Herault J. et~al. (2011) \emph{Physical Review E}, \emph{84}, 3, 36321.

\bibitem[\protect\astroncite{\emph{{Hirose} and
  {Turner}}}{2011}]{hirose&turner11}
{Hirose} S. and {Turner} N.~J. (2011) \emph{\apjl}, \emph{732}, L30.

\bibitem[\protect\astroncite{\emph{{Hopkins} and
  {Christiansen}}}{2013}]{HopChris13}
{Hopkins} P.~F. and {Christiansen} J.~L. (2013) \emph{\apj}, \emph{776}, 48.

\bibitem[\protect\astroncite{\emph{{Horne}}}{1995}]{1995A&A...297..273H}
{Horne} K. (1995) \emph{\aap}, \emph{297}, 273.

\bibitem[\protect\astroncite{\emph{{Hughes} and
  {Armitage}}}{2010}]{2010ApJ...719.1633H}
{Hughes} A.~L.~H. and {Armitage} P.~J. (2010) \emph{\apj}, \emph{719}, 1633.

\bibitem[\protect\astroncite{\emph{{Hughes}
  et~al.}}{2009}]{2009ApJ...704.1204H}
{Hughes} A.~M. et~al. (2009) \emph{\apj}, \emph{704}, 1204.

\bibitem[\protect\astroncite{\emph{{Hughes}
  et~al.}}{2011}]{2011ApJ...727...85H}
{Hughes} A.~M. et~al. (2011) \emph{\apj}, \emph{727}, 85.

\bibitem[\protect\astroncite{\emph{{Hughes}
  et~al.}}{2013}]{2013AJ....145..115H}
{Hughes} A.~M. et~al. (2013) \emph{\aj}, \emph{145}, 115.

\bibitem[\protect\astroncite{\emph{{Igea} and
  {Glassgold}}}{1999}]{1999apj...518..848i}
{Igea} J. and {Glassgold} A.~E. (1999) \emph{\apj}, \emph{518}, 848.

\bibitem[\protect\astroncite{\emph{{Ilee} et~al.}}{2011}]{2011MNRAS.417.2950I}
{Ilee} J.~D. et~al. (2011) \emph{\mnras}, \emph{417}, 2950.

\bibitem[\protect\astroncite{\emph{{Ilgner} and
  {Nelson}}}{2006{\natexlab{a}}}]{2006A&A...445..205I}
{Ilgner} M. and {Nelson} R.~P. (2006{\natexlab{a}}) \emph{\aap}, \emph{445},
  205.

\bibitem[\protect\astroncite{\emph{{Ilgner} and
  {Nelson}}}{2006{\natexlab{b}}}]{2006a&a...445..223i}
{Ilgner} M. and {Nelson} R.~P. (2006{\natexlab{b}}) \emph{\aap}, \emph{445},
  223.

\bibitem[\protect\astroncite{\emph{{Ilgner} and
  {Nelson}}}{2008}]{ilgner&nelson08}
{Ilgner} M. and {Nelson} R.~P. (2008) \emph{\aap}, \emph{483}, 815.

\bibitem[\protect\astroncite{\emph{{Inaba} and
  {Barge}}}{2006}]{2006ApJ...649..415I}
{Inaba} S. and {Barge} P. (2006) \emph{\apj}, \emph{649}, 415.

\bibitem[\protect\astroncite{\emph{{Isella}
  et~al.}}{2009}]{2009ApJ...701..260I}
{Isella} A. et~al. (2009) \emph{\apj}, \emph{701}, 260.

\bibitem[\protect\astroncite{\emph{{Ji} et~al.}}{2006}]{Ji06}
{Ji} H. et~al. (2006) \emph{\nat}, \emph{444}, 343.

\bibitem[\protect\astroncite{\emph{{Jin}}}{1996}]{1996ApJ...457..798J}
{Jin} L. (1996) \emph{\apj}, \emph{457}, 798.

\bibitem[\protect\astroncite{\emph{{Johansen} and
  {Klahr}}}{2005}]{2005ApJ...634.1353J}
{Johansen} A. and {Klahr} H. (2005) \emph{\apj}, \emph{634}, 1353.

\bibitem[\protect\astroncite{\emph{{Johansen}
  et~al.}}{2006{\natexlab{a}}}]{2006ApJ...643.1219J}
{Johansen} A. et~al. (2006{\natexlab{a}}) \emph{\apj}, \emph{643}, 1219.

\bibitem[\protect\astroncite{\emph{{Johansen}
  et~al.}}{2006{\natexlab{b}}}]{2006MNRAS.370L..71J}
{Johansen} A. et~al. (2006{\natexlab{b}}) \emph{\mnras}, \emph{370}, L71.

\bibitem[\protect\astroncite{\emph{{Johansen} et~al.}}{2009}]{johansenetal09}
{Johansen} A. et~al. (2009) \emph{\apj}, \emph{697}, 1269.

\bibitem[\protect\astroncite{\emph{{Johnson} and {Gammie}}}{2003}]{JohnGam03}
{Johnson} B.~M. and {Gammie} C.~F. (2003) \emph{\apj}, \emph{597}, 131.

\bibitem[\protect\astroncite{\emph{{Johnson} and {Gammie}}}{2005}]{Johnson05}
{Johnson} B.~M. and {Gammie} C.~F. (2005) \emph{\apj}, \emph{626}, 978.

\bibitem[\protect\astroncite{\emph{{Johnson} and {Gammie}}}{2006}]{Johnson06}
{Johnson} B.~M. and {Gammie} C.~F. (2006) \emph{\apj}, \emph{636}, 63.

\bibitem[\protect\astroncite{\emph{{Johnson} et~al.}}{2006}]{JohnGood06}
{Johnson} E.~T. et~al. (2006) \emph{\apj}, \emph{647}, 1413.

\bibitem[\protect\astroncite{\emph{{Juh{\'a}sz}
  et~al.}}{2012}]{2012ApJ...744..118J}
{Juh{\'a}sz} A. et~al. (2012) \emph{\apj}, \emph{744}, 118.

\bibitem[\protect\astroncite{\emph{{Julian} and {Toomre}}}{1966}]{JulToom66}
{Julian} W.~H. and {Toomre} A. (1966) \emph{\apj}, \emph{146}, 810.

\bibitem[\protect\astroncite{\emph{{Kemper}
  et~al.}}{2005}]{2005ApJ...633..534K}
{Kemper} F. et~al. (2005) \emph{\apj}, \emph{633}, 534.

\bibitem[\protect\astroncite{\emph{{Kim} and
  {Ostriker}}}{2000}]{2000ApJ...540..372K}
{Kim} W.-T. and {Ostriker} E.~C. (2000) \emph{\apj}, \emph{540}, 372.

\bibitem[\protect\astroncite{\emph{{Kitchatinov} and
  {R{\"u}diger}}}{2010}]{2010A&A...513L...1K}
{Kitchatinov} L.~L. and {R{\"u}diger} G. (2010) \emph{\aap}, \emph{513}, L1.

\bibitem[\protect\astroncite{\emph{{Klahr}}}{2004}]{Klahr04}
{Klahr} H. (2004) \emph{\apj}, \emph{606}, 1070.

\bibitem[\protect\astroncite{\emph{{Klahr}}}{2007}]{Klahr07}
{Klahr} H. (2007) in: \emph{IAU Symposium}, vol. 239, (edited by F.~{Kupka},
  I.~{Roxburgh}, and K.~L. {Chan}), pp. 405--416.

\bibitem[\protect\astroncite{\emph{{Klahr} and
  {Bodenheimer}}}{2006}]{KlahrBodenheimer06}
{Klahr} H. and {Bodenheimer} P. (2006) \emph{\apj}, \emph{639}, 432.

\bibitem[\protect\astroncite{\emph{{Klahr} and {Bodenheimer}}}{2003}]{kb03}
{Klahr} H.~H. and {Bodenheimer} P. (2003) \emph{\apj}, \emph{582}, 869.

\bibitem[\protect\astroncite{\emph{K{\"o}nigl et~al.}}{2010}]{ksw10}
K{\"o}nigl A. et~al. (2010) \emph{\mnras}, \emph{401}, 1, 479.

\bibitem[\protect\astroncite{\emph{{K{\"o}nigl}
  et~al.}}{2010}]{2010MNRAS.401..479K}
{K{\"o}nigl} A. et~al. (2010) \emph{\mnras}, \emph{401}, 479.

\bibitem[\protect\astroncite{\emph{{Kratter} et~al.}}{2010}]{Kratter10}
{Kratter} K.~M. et~al. (2010) \emph{\apj}, \emph{708}, 1585.

\bibitem[\protect\astroncite{\emph{{Krumholz} et~al.}}{2007}]{Krum07}
{Krumholz} M.~R. et~al. (2007) \emph{\apj}, \emph{656}, 959.

\bibitem[\protect\astroncite{\emph{{Kunz} and
  {Balbus}}}{2004}]{2004MNRAS.348..355K}
{Kunz} M.~W. and {Balbus} S.~A. (2004) \emph{\mnras}, \emph{348}, 355.

\bibitem[\protect\astroncite{\emph{{Kunz} and
  {Lesur}}}{2013}]{2013MNRAS.434.2295K}
{Kunz} M.~W. and {Lesur} G. (2013) \emph{\mnras}, \emph{434}, 2295.

\bibitem[\protect\astroncite{\emph{{Lathrop} et~al.}}{1992}]{Lathrop92}
{Lathrop} D.~P. et~al. (1992) \emph{Physical Review Letters}, \emph{68}, 1515.

\bibitem[\protect\astroncite{\emph{{Latter} et~al.}}{2010}]{lfg10}
{Latter} H.~N. et~al. (2010) \emph{\mnras}, \emph{406}, 848.

\bibitem[\protect\astroncite{\emph{{Lee} et~al.}}{2010}]{2010ApJ...718.1367L}
{Lee} A.~T. et~al. (2010) \emph{\apj}, \emph{718}, 1367.

\bibitem[\protect\astroncite{\emph{{Lesur} and {Longaretti}}}{2005}]{ll05}
{Lesur} G. and {Longaretti} P.-Y. (2005) \emph{\aap}, \emph{444}, 25.

\bibitem[\protect\astroncite{\emph{{Lesur} and
  {Longaretti}}}{2007}]{lesur&longaretti07}
{Lesur} G. and {Longaretti} P.-Y. (2007) \emph{\mnras}, \emph{378}, 1471.

\bibitem[\protect\astroncite{\emph{{Lesur} and
  {Longaretti}}}{2009}]{2009A&A...504..309L}
{Lesur} G. and {Longaretti} P.-Y. (2009) \emph{\aap}, \emph{504}, 309.

\bibitem[\protect\astroncite{\emph{{Lesur} and {Ogilvie}}}{2010}]{lo10}
{Lesur} G. and {Ogilvie} G.~I. (2010) \emph{\mnras}, \emph{404}, L64.

\bibitem[\protect\astroncite{\emph{{Lesur} and {Papaloizou}}}{2009}]{lp09}
{Lesur} G. and {Papaloizou} J.~C.~B. (2009) \emph{\aap}, \emph{498}, 1.

\bibitem[\protect\astroncite{\emph{{Lesur} and {Papaloizou}}}{2010}]{lp10}
{Lesur} G. and {Papaloizou} J.~C.~B. (2010) \emph{\aap}, \emph{513}, A60+.

\bibitem[\protect\astroncite{\emph{{Lesur} et~al.}}{2013}]{lfo13}
{Lesur} G. et~al. (2013) \emph{\aap}, \emph{550}, A61.

\bibitem[\protect\astroncite{\emph{{Li} et~al.}}{2000}]{Li00}
{Li} H. et~al. (2000) \emph{\apj}, \emph{533}, 1023.

\bibitem[\protect\astroncite{\emph{{Li} et~al.}}{2001}]{Li01}
{Li} H. et~al. (2001) \emph{\apj}, \emph{551}, 874.

\bibitem[\protect\astroncite{\emph{{Lin} and
  {Papaloizou}}}{1993}]{1993prpl.conf..749L}
{Lin} D.~N.~C. and {Papaloizou} J.~C.~B. (1993) in: \emph{Protostars and
  Planets III}, (edited by E.~H. {Levy} and J.~I. {Lunine}), pp. 749--835.

\bibitem[\protect\astroncite{\emph{{Lisse} et~al.}}{2006}]{2006Sci...313..635L}
{Lisse} C.~M. et~al. (2006) \emph{Science}, \emph{313}, 635.

\bibitem[\protect\astroncite{\emph{{Lodato} and {Rice}}}{2004}]{LodatoRice04}
{Lodato} G. and {Rice} W.~K.~M. (2004) \emph{\mnras}, \emph{351}, 630.

\bibitem[\protect\astroncite{\emph{{Lodato} and {Rice}}}{2005}]{LodatoRice05}
{Lodato} G. and {Rice} W.~K.~M. (2005) \emph{\mnras}, \emph{358}, 1489.

\bibitem[\protect\astroncite{\emph{{Longaretti} and
  {Lesur}}}{2010}]{longaretti&lesur10}
{Longaretti} P.-Y. and {Lesur} G. (2010) \emph{\aap}, \emph{516}, A51.

\bibitem[\protect\astroncite{\emph{{Lovelace} et~al.}}{1999}]{Colgate99}
{Lovelace} R.~V.~E. et~al. (1999) \emph{\apj}, \emph{513}, 805.

\bibitem[\protect\astroncite{\emph{{Lubow} and {Spruit}}}{1995}]{ls95}
{Lubow} S.~H. and {Spruit} H.~C. (1995) \emph{\apj}, \emph{445}, 337.

\bibitem[\protect\astroncite{\emph{{Lubow} et~al.}}{1994{\natexlab{a}}}]{lpp94}
{Lubow} S.~H. et~al. (1994{\natexlab{a}}) \emph{\mnras}, \emph{267}, 235.

\bibitem[\protect\astroncite{\emph{{Lubow}
  et~al.}}{1994{\natexlab{b}}}]{lpp94b}
{Lubow} S.~H. et~al. (1994{\natexlab{b}}) \emph{\mnras}, \emph{268}, 1010.

\bibitem[\protect\astroncite{\emph{{Luhman}
  et~al.}}{2008}]{2008ApJ...675.1375L}
{Luhman} K.~L. et~al. (2008) \emph{\apj}, \emph{675}, 1375.

\bibitem[\protect\astroncite{\emph{{Luhman}
  et~al.}}{2010}]{2010ApJS..186..111L}
{Luhman} K.~L. et~al. (2010) \emph{\apjs}, \emph{186}, 111.

\bibitem[\protect\astroncite{\emph{{Lyra} and {Klahr}}}{2011}]{lk11}
{Lyra} W. and {Klahr} H. (2011) \emph{\aap}, \emph{527}, A138.

\bibitem[\protect\astroncite{\emph{{Lyra} and
  {Lin}}}{2013}]{2013ApJ...775...17L}
{Lyra} W. and {Lin} M.-K. (2013) \emph{\apj}, \emph{775}, 17.

\bibitem[\protect\astroncite{\emph{{Lyra} and {Mac Low}}}{2012}]{LyraMacMlow12}
{Lyra} W. and {Mac Low} M.-M. (2012) \emph{\apj}, \emph{756}, 62.

\bibitem[\protect\astroncite{\emph{{Lyra} et~al.}}{2008}]{2008A&A...491L..41L}
{Lyra} W. et~al. (2008) \emph{\aap}, \emph{491}, L41.

\bibitem[\protect\astroncite{\emph{{Lyra} et~al.}}{2009}]{2009A&A...497..869L}
{Lyra} W. et~al. (2009) \emph{\aap}, \emph{497}, 869.

\bibitem[\protect\astroncite{\emph{{Mac Low}
  et~al.}}{1995}]{1995ApJ...442..726M}
{Mac Low} M.-M. et~al. (1995) \emph{\apj}, \emph{442}, 726.

\bibitem[\protect\astroncite{\emph{{Mamatsashvili} and
  {Rice}}}{2010}]{MamRice10}
{Mamatsashvili} G.~R. and {Rice} W.~K.~M. (2010) \emph{\mnras}, \emph{406},
  2050.

\bibitem[\protect\astroncite{\emph{{Marcus}
  et~al.}}{2013}]{2013PhRvL.111h4501M}
{Marcus} P.~S. et~al. (2013) \emph{Physical Review Letters}, \emph{111}, 8,
  084501.

\bibitem[\protect\astroncite{\emph{{Matzner} and {Levin}}}{2005}]{matzlev05}
{Matzner} C.~D. and {Levin} Y. (2005) \emph{\apj}, \emph{628}, 817.

\bibitem[\protect\astroncite{\emph{{Meheut} et~al.}}{2010}]{Meheut10}
{Meheut} H. et~al. (2010) \emph{\aap}, \emph{516}, A31.

\bibitem[\protect\astroncite{\emph{{Meheut}
  et~al.}}{2012{\natexlab{a}}}]{2012A&A...545A.134M}
{Meheut} H. et~al. (2012{\natexlab{a}}) \emph{\aap}, \emph{545}, A134.

\bibitem[\protect\astroncite{\emph{{Meheut}
  et~al.}}{2012{\natexlab{b}}}]{Meheut12a}
{Meheut} H. et~al. (2012{\natexlab{b}}) \emph{\aap}, \emph{542}, A9.

\bibitem[\protect\astroncite{\emph{{Meheut}
  et~al.}}{2012{\natexlab{c}}}]{Meheut12b}
{Meheut} H. et~al. (2012{\natexlab{c}}) \emph{\mnras}, \emph{422}, 2399.

\bibitem[\protect\astroncite{\emph{{Meru} and {Bate}}}{2011}]{MeruBate11}
{Meru} F. and {Bate} M.~R. (2011) \emph{\mnras}, \emph{411}, L1.

\bibitem[\protect\astroncite{\emph{{Meru} and {Bate}}}{2012}]{MeruBate12}
{Meru} F. and {Bate} M.~R. (2012) \emph{\mnras}, \emph{427}, 2022.

\bibitem[\protect\astroncite{\emph{{Meschiari} and {Laughlin}}}{2008}]{Mesch08}
{Meschiari} S. and {Laughlin} G. (2008) \emph{\apjl}, \emph{679}, L135.

\bibitem[\protect\astroncite{\emph{{Michael} et~al.}}{2011}]{Michael11}
{Michael} S. et~al. (2011) \emph{\apjl}, \emph{737}, L42.

\bibitem[\protect\astroncite{\emph{{Michael} et~al.}}{2012}]{Michael12}
{Michael} S. et~al. (2012) \emph{\apj}, \emph{746}, 98.

\bibitem[\protect\astroncite{\emph{{Mignone} et~al.}}{2007}]{mignoneetal07}
{Mignone} A. et~al. (2007) \emph{\apjs}, \emph{170}, 228.

\bibitem[\protect\astroncite{\emph{{Miller} and
  {Stone}}}{2000}]{2000apj...534..398m}
{Miller} K.~A. and {Stone} J.~M. (2000) \emph{\apj}, \emph{534}, 398.

\bibitem[\protect\astroncite{\emph{{Min} et~al.}}{2005}]{2005Icar..179..158M}
{Min} M. et~al. (2005) \emph{\icarus}, \emph{179}, 158.

\bibitem[\protect\astroncite{\emph{{Mohanty} et~al.}}{2013}]{mohantyetal13}
{Mohanty} S. et~al. (2013) \emph{\apj}, \emph{764}, 65.

\bibitem[\protect\astroncite{\emph{{Moll}}}{2012}]{m12}
{Moll} R. (2012) \emph{\aap}, \emph{548}, A76.

\bibitem[\protect\astroncite{\emph{{Morales-Calder{\'o}n}
  et~al.}}{2011}]{2011ApJ...733...50M}
{Morales-Calder{\'o}n} M. et~al. (2011) \emph{\apj}, \emph{733}, 50.

\bibitem[\protect\astroncite{\emph{{Mulders} and
  {Dominik}}}{2012}]{2012A&A...539A...9M}
{Mulders} G.~D. and {Dominik} C. (2012) \emph{\aap}, \emph{539}, A9.

\bibitem[\protect\astroncite{\emph{{M{\"u}ller} et~al.}}{2012}]{Muller12}
{M{\"u}ller} T.~W.~A. et~al. (2012) \emph{\aap}, \emph{541}, A123.

\bibitem[\protect\astroncite{\emph{{Murphy} et~al.}}{2010}]{mfz10}
{Murphy} G.~C. et~al. (2010) \emph{\aap}, \emph{512}, A82.

\bibitem[\protect\astroncite{\emph{{Muzerolle}
  et~al.}}{2009}]{2009ApJ...704L..15M}
{Muzerolle} J. et~al. (2009) \emph{\apjl}, \emph{704}, L15.

\bibitem[\protect\astroncite{\emph{{Najita}
  et~al.}}{2007}]{2007prpl.conf..507N}
{Najita} J.~R. et~al. (2007) \emph{Protostars and Planets V}, pp. 507--522.

\bibitem[\protect\astroncite{\emph{{Najita}
  et~al.}}{2009}]{2009ApJ...691..738N}
{Najita} J.~R. et~al. (2009) \emph{\apj}, \emph{691}, 738.

\bibitem[\protect\astroncite{\emph{{Najita}
  et~al.}}{2013}]{2013ApJ...766..134N}
{Najita} J.~R. et~al. (2013) \emph{\apj}, \emph{766}, 134.

\bibitem[\protect\astroncite{\emph{{Nelson} and
  {Papaloizou}}}{2004}]{2004MNRAS.350..849N}
{Nelson} R.~P. and {Papaloizou} J.~C.~B. (2004) \emph{\mnras}, \emph{350}, 849.

\bibitem[\protect\astroncite{\emph{{Nelson}
  et~al.}}{2013}]{2013MNRAS.435.2610N}
{Nelson} R.~P. et~al. (2013) \emph{\mnras}, \emph{435}, 2610.

\bibitem[\protect\astroncite{\emph{{Ogilvie}}}{2012}]{o12}
{Ogilvie} G.~I. (2012) \emph{\mnras}, \emph{423}, 1318.

\bibitem[\protect\astroncite{\emph{{Oishi} and {Mac
  Low}}}{2009}]{2009ApJ...704.1239O}
{Oishi} J.~S. and {Mac Low} M.-M. (2009) \emph{\apj}, \emph{704}, 1239.

\bibitem[\protect\astroncite{\emph{{Okuzumi}}}{2009}]{2009ApJ...698.1122O}
{Okuzumi} S. (2009) \emph{\apj}, \emph{698}, 1122.

\bibitem[\protect\astroncite{\emph{{Okuzumi} and
  {Hirose}}}{2011}]{okuzumi&hirose11}
{Okuzumi} S. and {Hirose} S. (2011) \emph{\apj}, \emph{742}, 65.

\bibitem[\protect\astroncite{\emph{{Owen} et~al.}}{2010}]{2010MNRAS.401.1415O}
{Owen} J.~E. et~al. (2010) \emph{\mnras}, \emph{401}, 1415.

\bibitem[\protect\astroncite{\emph{{Paardekooper}}}{2012}]{Pard12}
{Paardekooper} S.-J. (2012) \emph{\mnras}, \emph{421}, 3286.

\bibitem[\protect\astroncite{\emph{{Paardekooper}
  et~al.}}{2010}]{PaardekooperPapaloizou10}
{Paardekooper} S.-J. et~al. (2010) \emph{\apj}, \emph{725}, 146.

\bibitem[\protect\astroncite{\emph{{Paardekooper} et~al.}}{2011}]{Pard11}
{Paardekooper} S.-J. et~al. (2011) \emph{\mnras}, \emph{416}, L65.

\bibitem[\protect\astroncite{\emph{{Paczynski}}}{1978}]{Pac78}
{Paczynski} B. (1978) \emph{\actaa}, \emph{28}, 91.

\bibitem[\protect\astroncite{\emph{{Pandey} and
  {Wardle}}}{2012}]{2012MNRAS.423..222P}
{Pandey} B.~P. and {Wardle} M. (2012) \emph{\mnras}, \emph{423}, 222.

\bibitem[\protect\astroncite{\emph{{Panoglou}
  et~al.}}{2012}]{2012A&A...538A...2P}
{Panoglou} D. et~al. (2012) \emph{\aap}, \emph{538}, A2.

\bibitem[\protect\astroncite{\emph{{Paoletti} et~al.}}{2012}]{Paoletti12}
{Paoletti} M.~S. et~al. (2012) \emph{\aap}, \emph{547}, A64.

\bibitem[\protect\astroncite{\emph{{Perez-Becker} and
  {Chiang}}}{2011{\natexlab{a}}}]{2011ApJ...735....8P}
{Perez-Becker} D. and {Chiang} E. (2011{\natexlab{a}}) \emph{\apj}, \emph{735},
  8.

\bibitem[\protect\astroncite{\emph{{Perez-Becker} and
  {Chiang}}}{2011{\natexlab{b}}}]{2011ApJ...727....2P}
{Perez-Becker} D. and {Chiang} E. (2011{\natexlab{b}}) \emph{\apj}, \emph{727},
  2.

\bibitem[\protect\astroncite{\emph{{Petersen}
  et~al.}}{2007{\natexlab{a}}}]{pjs07}
{Petersen} M.~R. et~al. (2007{\natexlab{a}}) \emph{\apj}, \emph{658}, 1236.

\bibitem[\protect\astroncite{\emph{{Petersen}
  et~al.}}{2007{\natexlab{b}}}]{Petersen07b}
{Petersen} M.~R. et~al. (2007{\natexlab{b}}) \emph{\apj}, \emph{658}, 1252.

\bibitem[\protect\astroncite{\emph{{Pinte} et~al.}}{2008}]{2008A&A...489..633P}
{Pinte} C. et~al. (2008) \emph{\aap}, \emph{489}, 633.

\bibitem[\protect\astroncite{\emph{{Pontoppidan}
  et~al.}}{2010}]{2010ApJ...720..887P}
{Pontoppidan} K.~M. et~al. (2010) \emph{\apj}, \emph{720}, 887.

\bibitem[\protect\astroncite{\emph{{Raettig} et~al.}}{2013}]{Raettig13}
{Raettig} N. et~al. (2013) \emph{\apj}, \emph{765}, 115.

\bibitem[\protect\astroncite{\emph{{Rafikov}}}{2005}]{rafikov05}
{Rafikov} R.~R. (2005) \emph{\apjl}, \emph{621}, L69.

\bibitem[\protect\astroncite{\emph{{Rafikov}}}{2009}]{Rafikov09}
{Rafikov} R.~R. (2009) \emph{\apj}, \emph{704}, 281.

\bibitem[\protect\astroncite{\emph{{Rayleigh}}}{1917}]{Rayleigh17}
{Rayleigh} L. (1917) \emph{Royal Society of London Proceedings Series A},
  \emph{93}, 148.

\bibitem[\protect\astroncite{\emph{{Reg{\'a}ly}
  et~al.}}{2012}]{2012MNRAS.419.1701R}
{Reg{\'a}ly} Z. et~al. (2012) \emph{\mnras}, \emph{419}, 1701.

\bibitem[\protect\astroncite{\emph{{Rice} et~al.}}{2011}]{Rice11}
{Rice} W.~K.~M. et~al. (2011) \emph{\mnras}, \emph{418}, 1356.

\bibitem[\protect\astroncite{\emph{{Richard} and {Zahn}}}{1999}]{Richard99}
{Richard} D. and {Zahn} J.-P. (1999) \emph{\aap}, \emph{347}, 734.

\bibitem[\protect\astroncite{\emph{{Rigliaco}
  et~al.}}{2013}]{2013ApJ...772...60R}
{Rigliaco} E. et~al. (2013) \emph{\apj}, \emph{772}, 60.

\bibitem[\protect\astroncite{\emph{{Romanova}
  et~al.}}{2013}]{2013MNRAS.430..699R}
{Romanova} M.~M. et~al. (2013) \emph{\mnras}, \emph{430}, 699.

\bibitem[\protect\astroncite{\emph{{R{\"u}diger}
  et~al.}}{2002}]{2002A&A...391..781R}
{R{\"u}diger} G. et~al. (2002) \emph{\aap}, \emph{391}, 781.

\bibitem[\protect\astroncite{\emph{{Ruge} et~al.}}{2013}]{ruge13}
{Ruge} J.~P. et~al. (2013) \emph{\aap}, \emph{549}, A97.

\bibitem[\protect\astroncite{\emph{{Ruzmaikina}}}{1982}]{1982MitAG..57...49R}
{Ruzmaikina} T.~V. (1982) \emph{Mitteilungen der Astronomischen Gesellschaft
  Hamburg}, \emph{57}, 45.

\bibitem[\protect\astroncite{\emph{{Salmeron} and
  {Wardle}}}{2008}]{2008MNRAS.388.1223S}
{Salmeron} R. and {Wardle} M. (2008) \emph{\mnras}, \emph{388}, 1223.

\bibitem[\protect\astroncite{\emph{Salmeron et~al.}}{2011}]{skw11}
Salmeron R. et~al. (2011) \emph{\mnras}, \emph{412}, 2, 1162.

\bibitem[\protect\astroncite{\emph{{Salyk} et~al.}}{2008}]{2008ApJ...676L..49S}
{Salyk} C. et~al. (2008) \emph{\apjl}, \emph{676}, L49.

\bibitem[\protect\astroncite{\emph{{Salyk} et~al.}}{2011}]{2011ApJ...731..130S}
{Salyk} C. et~al. (2011) \emph{\apj}, \emph{731}, 130.

\bibitem[\protect\astroncite{\emph{{Sano} and
  {Inutsuka}}}{2001}]{2001ApJ...561L.179S}
{Sano} T. and {Inutsuka} S.-i. (2001) \emph{\apjl}, \emph{561}, L179.

\bibitem[\protect\astroncite{\emph{{Sano} and
  {Miyama}}}{1999}]{1999ApJ...515..776S}
{Sano} T. and {Miyama} S.~M. (1999) \emph{\apj}, \emph{515}, 776.

\bibitem[\protect\astroncite{\emph{{Sano} and
  {Stone}}}{2002{\natexlab{a}}}]{2002apj...570..314s}
{Sano} T. and {Stone} J.~M. (2002{\natexlab{a}}) \emph{\apj}, \emph{570}, 314.

\bibitem[\protect\astroncite{\emph{{Sano} and
  {Stone}}}{2002{\natexlab{b}}}]{2002apj...577..534s}
{Sano} T. and {Stone} J.~M. (2002{\natexlab{b}}) \emph{\apj}, \emph{577}, 534.

\bibitem[\protect\astroncite{\emph{{Sano} et~al.}}{2000}]{2000apj...543..486s}
{Sano} T. et~al. (2000) \emph{\apj}, \emph{543}, 486.

\bibitem[\protect\astroncite{\emph{{Sargent}
  et~al.}}{2009}]{2009ApJS..182..477S}
{Sargent} B.~A. et~al. (2009) \emph{\apjs}, \emph{182}, 477.

\bibitem[\protect\astroncite{\emph{{Schartman} et~al.}}{2012}]{Schartman12}
{Schartman} E. et~al. (2012) \emph{\aap}, \emph{543}, A94.

\bibitem[\protect\astroncite{\emph{{Schultz-Grunow}}}{1959}]{Schultz-Grunow59}
{Schultz-Grunow} F. (1959) \emph{Z. Angewandte Mathematik und Mechanik},
  \emph{39}, 101.

\bibitem[\protect\astroncite{\emph{{Semenov} and
  {Wiebe}}}{2011}]{2011apjs..196...25s}
{Semenov} D. and {Wiebe} D. (2011) \emph{\apjs}, \emph{196}, 25.

\bibitem[\protect\astroncite{\emph{{Shakura} and
  {Sunyaev}}}{1973}]{1973a&a....24..337s}
{Shakura} N.~I. and {Sunyaev} R.~A. (1973) \emph{\aap}, \emph{24}, 337.

\bibitem[\protect\astroncite{\emph{{Shen} et~al.}}{2006}]{2006ApJ...653..513S}
{Shen} Y. et~al. (2006) \emph{\apj}, \emph{653}, 513.

\bibitem[\protect\astroncite{\emph{{Shi} et~al.}}{2010}]{shietal10}
{Shi} J. et~al. (2010) \emph{\apj}, \emph{708}, 1716.

\bibitem[\protect\astroncite{\emph{{Shlosman} and
  {Begelman}}}{1987}]{ShlosBeg87}
{Shlosman} I. and {Begelman} M.~C. (1987) \emph{\nat}, \emph{329}, 810.

\bibitem[\protect\astroncite{\emph{{Simon} and
  {Hawley}}}{2009}]{simon&hawley09}
{Simon} J.~B. and {Hawley} J.~F. (2009) \emph{\apj}, \emph{707}, 833.

\bibitem[\protect\astroncite{\emph{{Simon} et~al.}}{2009}]{simon&hawley09a}
{Simon} J.~B. et~al. (2009) \emph{\apj}, \emph{690}, 974.

\bibitem[\protect\astroncite{\emph{{Simon}
  et~al.}}{2011{\natexlab{a}}}]{simonetal11}
{Simon} J.~B. et~al. (2011{\natexlab{a}}) \emph{\apj}, \emph{730}, 94.

\bibitem[\protect\astroncite{\emph{{Simon}
  et~al.}}{2011{\natexlab{b}}}]{2011ApJ...743...17S}
{Simon} J.~B. et~al. (2011{\natexlab{b}}) \emph{\apj}, \emph{743}, 17.

\bibitem[\protect\astroncite{\emph{{Simon} et~al.}}{2012}]{simonetal12}
{Simon} J.~B. et~al. (2012) \emph{\mnras}, \emph{422}, 2685.

\bibitem[\protect\astroncite{\emph{{Simon}
  et~al.}}{2013{\natexlab{a}}}]{simonetal13}
{Simon} J.~B. et~al. (2013{\natexlab{a}}) \emph{\apj}, \emph{764}, 66.

\bibitem[\protect\astroncite{\emph{{Simon} et~al.}}{2013{\natexlab{b}}}]{sba13}
{Simon} J.~B. et~al. (2013{\natexlab{b}}) \emph{\apj}, \emph{775}, 73.

\bibitem[\protect\astroncite{\emph{{Sorathia} et~al.}}{2010}]{sorathiaetal10}
{Sorathia} K.~A. et~al. (2010) \emph{\apj}, \emph{712}, 1241.

\bibitem[\protect\astroncite{\emph{{Sorathia} et~al.}}{2012}]{sorathiaetal12}
{Sorathia} K.~A. et~al. (2012) \emph{\apj}, \emph{749}, 189.

\bibitem[\protect\astroncite{\emph{{Spruit}}}{1996}]{s96}
{Spruit} H.~C. (1996) in: \emph{NATO ASIC Proc. 477: Evolutionary Processes in
  Binary Stars}, (edited by R.~A.~M.~J. {Wijers}, M.~B. {Davies}, and C.~A.
  {Tout}), pp. 249--286.

\bibitem[\protect\astroncite{\emph{{Steiman-Cameron}
  et~al.}}{2013}]{SteimanCam13}
{Steiman-Cameron} T.~Y. et~al. (2013) \emph{\apj}, \emph{768}, 192.

\bibitem[\protect\astroncite{\emph{{Stone} and
  {Norman}}}{1992}]{1992apjs...80..753s}
{Stone} J.~M. and {Norman} M.~L. (1992) \emph{\apjs}, \emph{80}, 753.

\bibitem[\protect\astroncite{\emph{{Stone} et~al.}}{2008}]{stoneetal08}
{Stone} J.~M. et~al. (2008) \emph{\apjs}, \emph{178}, 137.

\bibitem[\protect\astroncite{\emph{{Suzuki} and {Inutsuka}}}{2009}]{si09}
{Suzuki} T.~K. and {Inutsuka} S.-i. (2009) \emph{\apjl}, \emph{691}, L49.

\bibitem[\protect\astroncite{\emph{Suzuki et~al.}}{2010}]{smi10}
Suzuki T.~K. et~al. (2010) \emph{\apj}, \emph{718}, 2, 1289.

\bibitem[\protect\astroncite{\emph{{Tanga} et~al.}}{1996}]{Tanga96}
{Tanga} P. et~al. (1996) \emph{\icarus}, \emph{121}, 158.

\bibitem[\protect\astroncite{\emph{{Tassoul}}}{2007}]{Tassoul07}
{Tassoul} J.-L. (2007) \emph{{Stellar Rotation}}, Cambridge University Press; 1
  edition (July 2, 2007).

\bibitem[\protect\astroncite{\emph{{Taylor}}}{1923}]{Taylor23}
{Taylor} G.~I. (1923) \emph{Royal Society of London Philosophical Transactions
  Series A}, \emph{223}, 289.

\bibitem[\protect\astroncite{\emph{{Taylor}}}{1936}]{Taylor36}
{Taylor} G.~I. (1936) \emph{Royal Society of London Proceedings Series A},
  \emph{157}, 546.

\bibitem[\protect\astroncite{\emph{{Terquem} and
  {Papaloizou}}}{1996}]{1996MNRAS.279..767T}
{Terquem} C. and {Papaloizou} J.~C.~B. (1996) \emph{\mnras}, \emph{279}, 767.

\bibitem[\protect\astroncite{\emph{{Turner} and
  {Drake}}}{2009}]{2009apj...703.2152t}
{Turner} N.~J. and {Drake} J.~F. (2009) \emph{\apj}, \emph{703}, 2152.

\bibitem[\protect\astroncite{\emph{{Turner} and
  {Sano}}}{2008}]{2008ApJ...679L.131T}
{Turner} N.~J. and {Sano} T. (2008) \emph{\apjl}, \emph{679}, L131.

\bibitem[\protect\astroncite{\emph{{Turner}
  et~al.}}{2006}]{2006apj...639.1218t}
{Turner} N.~J. et~al. (2006) \emph{\apj}, \emph{639}, 1218.

\bibitem[\protect\astroncite{\emph{{Turner}
  et~al.}}{2007}]{2007ApJ...659..729T}
{Turner} N.~J. et~al. (2007) \emph{\apj}, \emph{659}, 729.

\bibitem[\protect\astroncite{\emph{{Turner}
  et~al.}}{2010}]{2010ApJ...708..188T}
{Turner} N.~J. et~al. (2010) \emph{\apj}, \emph{708}, 188.

\bibitem[\protect\astroncite{\emph{{Tzeferacos} et~al.}}{2009}]{tfm09}
{Tzeferacos} P. et~al. (2009) \emph{\mnras}, \emph{400}, 820.

\bibitem[\protect\astroncite{\emph{{Tzeferacos} et~al.}}{2013}]{tfm13}
{Tzeferacos} P. et~al. (2013) \emph{\mnras}, \emph{428}, 3151.

\bibitem[\protect\astroncite{\emph{{Umebayashi} and
  {Nakano}}}{1981}]{1981PASJ...33..617U}
{Umebayashi} T. and {Nakano} T. (1981) \emph{\pasj}, \emph{33}, 617.

\bibitem[\protect\astroncite{\emph{Umebayashi and Nakano}}{1990}]{UN90}
Umebayashi T. and Nakano T. (1990) \emph{MNRAS}, \emph{243}, 103.

\bibitem[\protect\astroncite{\emph{{Umebayashi} and
  {Nakano}}}{2009}]{2009apj...690...69u}
{Umebayashi} T. and {Nakano} T. (2009) \emph{\apj}, \emph{690}, 69.

\bibitem[\protect\astroncite{\emph{{Umebayashi}
  et~al.}}{2013}]{2013ApJ...764..104U}
{Umebayashi} T. et~al. (2013) \emph{\apj}, \emph{764}, 104.

\bibitem[\protect\astroncite{\emph{{Uribe} et~al.}}{2011}]{uribeetal11}
{Uribe} A.~L. et~al. (2011) \emph{\apj}, \emph{736}, 85.

\bibitem[\protect\astroncite{\emph{{Urpin}}}{2003}]{2003A&A...404..397U}
{Urpin} V. (2003) \emph{\aap}, \emph{404}, 397.

\bibitem[\protect\astroncite{\emph{{Urpin} and
  {Brandenburg}}}{1998}]{1998MNRAS.294..399U}
{Urpin} V. and {Brandenburg} A. (1998) \emph{\mnras}, \emph{294}, 399.

\bibitem[\protect\astroncite{\emph{{van der Marel}
  et~al.}}{2013}]{2013Sci...340.1199V}
{van der Marel} N. et~al. (2013) \emph{Science}, \emph{340}, 1199.

\bibitem[\protect\astroncite{\emph{{Varni{\`e}re} and
  {Tagger}}}{2006}]{Varniere06}
{Varni{\`e}re} P. and {Tagger} M. (2006) \emph{\aap}, \emph{446}, L13.

\bibitem[\protect\astroncite{\emph{Velikhov}}{1959}]{V59}
Velikhov E. (1959) \emph{J. Expt. Theor. Phys. (USSR)}, \emph{36}, 1398.

\bibitem[\protect\astroncite{\emph{{Vorobyov} and
  {Basu}}}{2007}]{2007MNRAS.381.1009V}
{Vorobyov} E.~I. and {Basu} S. (2007) \emph{\mnras}, \emph{381}, 1009.

\bibitem[\protect\astroncite{\emph{{Vorobyov} and {Basu}}}{2009}]{VoroBasu09}
{Vorobyov} E.~I. and {Basu} S. (2009) \emph{\mnras}, \emph{393}, 822.

\bibitem[\protect\astroncite{\emph{{Vorobyov} and {Basu}}}{2010}]{Voro10b}
{Vorobyov} E.~I. and {Basu} S. (2010) \emph{\apj}, \emph{719}, 1896.

\bibitem[\protect\astroncite{\emph{{Wardle}}}{1999}]{1999MNRAS.307..849W}
{Wardle} M. (1999) \emph{\mnras}, \emph{307}, 849.

\bibitem[\protect\astroncite{\emph{{Wardle}}}{2007}]{2007ap&ss.311...35w}
{Wardle} M. (2007) \emph{\apss}, \emph{311}, 35.

\bibitem[\protect\astroncite{\emph{{Wardle} and {K\"onigl}}}{1993}]{wk93}
{Wardle} M. and {K\"onigl} A. (1993) \emph{\apj}, \emph{410}, 218.

\bibitem[\protect\astroncite{\emph{{Wardle} and
  {Salmeron}}}{2012}]{2012MNRAS.422.2737W}
{Wardle} M. and {Salmeron} R. (2012) \emph{\mnras}, \emph{422}, 2737.

\bibitem[\protect\astroncite{\emph{{Watson}
  et~al.}}{2007}]{2007prpl.conf..523W}
{Watson} A.~M. et~al. (2007) \emph{Protostars and Planets V}, pp. 523--538.

\bibitem[\protect\astroncite{\emph{{Weidenschilling} and
  {Cuzzi}}}{1993}]{1993prpl.conf.1031W}
{Weidenschilling} S.~J. and {Cuzzi} J.~N. (1993) in: \emph{Protostars and
  Planets III}, (edited by E.~H. {Levy} and J.~I. {Lunine}), pp. 1031--1060.

\bibitem[\protect\astroncite{\emph{{Weiss} et~al.}}{2010}]{2010SSRv..152..341W}
{Weiss} B.~P. et~al. (2010) \emph{\ssr}, \emph{152}, 341.

\bibitem[\protect\astroncite{\emph{{Wendt}}}{1959}]{Wendt33}
{Wendt} F. (1959) \emph{Ingenieur-Archiv}, \emph{4}, 557.

\bibitem[\protect\astroncite{\emph{{Willacy}
  et~al.}}{2006}]{2006ApJ...644.1202W}
{Willacy} K. et~al. (2006) \emph{\apj}, \emph{644}, 1202.

\bibitem[\protect\astroncite{\emph{{Wolf} and {Klahr}}}{2002}]{WolfKLahr02}
{Wolf} S. and {Klahr} H. (2002) \emph{\apjl}, \emph{578}, L79.

\bibitem[\protect\astroncite{\emph{{Zanni} et~al.}}{2007}]{zfr07}
{Zanni} C. et~al. (2007) \emph{\aap}, \emph{469}, 811.

\bibitem[\protect\astroncite{\emph{{Ziegler}}}{2005}]{ziegler05}
{Ziegler} U. (2005) \emph{\aap}, \emph{435}, 385.

\bibitem[\protect\astroncite{\emph{{Ziegler}}}{2008}]{ziegler08}
{Ziegler} U. (2008) \emph{Computer Physics Communications}, \emph{179}, 227.

\bibitem[\protect\astroncite{\emph{{Zolensky}
  et~al.}}{2006}]{2006Sci...314.1735Z}
{Zolensky} M.~E. et~al. (2006) \emph{Science}, \emph{314}, 1735.

\bibitem[\protect\astroncite{\emph{{Zsom} et~al.}}{2010}]{2010a&a...513a..57z}
{Zsom} A. et~al. (2010) \emph{\aap}, \emph{513}, A57+.

\end{thebibliography}
\end{document}